\documentclass[sigchi]{acmart}
\AtBeginDocument{%
  }

\setcopyright{acmlicensed}
\copyrightyear{2026}
\acmYear{2026}
\acmDOI{XXXXXXX.XXXXXXX}
\acmConference[Conference acronym 'XX]{Make sure to enter the correct
  conference title from your rights confirmation email}{June 03--05,
  2018}{Woodstock, NY}
\acmISBN{978-1-4503-XXXX-X/2018/06}

\usepackage[most]{tcolorbox}
\usepackage{xcolor}
\usepackage{enumitem}

\usepackage{xspace}
\newcommand{\system}{LINK\xspace}

\usepackage{tabularx}
\usepackage{colortbl}
\usepackage{ragged2e}

\usepackage[most]{tcolorbox}

\definecolor{CommitmentBlue}{HTML}{315C73}
\definecolor{CommitmentLight}{HTML}{EAF1F4}

\definecolor{PatternLight}{HTML}{F3F6F7}
\definecolor{MutedText}{HTML}{58656B}

\definecolor{WhenGreen}{HTML}{3F6B52}
\definecolor{TensionOrange}{HTML}{8A5A2B}

\newcolumntype{Y}{>{\RaggedRight\arraybackslash}X}
\newcolumntype{P}[1]{>{\RaggedRight\arraybackslash}p{#1}}

\newtcolorbox{commitmentbox}{
    enhanced,
    breakable,
    colback=CommitmentLight,
    colframe=CommitmentBlue,
    boxrule=0pt,
    leftrule=2.5pt,
    arc=1mm,
    left=6pt,
    right=6pt,
    top=5pt,
    bottom=5pt,
    before skip=4pt,
    after skip=7pt
}

\newcommand{\designcommitment}[1]{%
\begin{commitmentbox}
    \textcolor{CommitmentBlue}{\textbf{\scriptsize DESIGN COMMITMENT}}\par
    \vspace{2pt}
    \textbf{#1}
\end{commitmentbox}
}

\newcommand{\patternname}[1]{%
    \textcolor{CommitmentBlue}{\textbf{#1}}
}

\newcommand{\mechanism}[2]{%
    \textbf{\textcolor{CommitmentBlue}{#1}}%
    \enspace$\rightarrow$\enspace #2%
}

\newcommand{\corpusexample}[1]{%
    \par\vspace{3pt}%
    \textcolor{MutedText}{%
        \textbf{\scriptsize EXAMPLES}\enspace #1%
    }%
}

\newcommand{\when}[1]{%
    \textcolor{WhenGreen}{%
        \(\checkmark\)\,\textbf{\scriptsize WHEN}%
    }%
    \enspace #1\par
}

\newcommand{\tension}[1]{%
    \vspace{3pt}%
    \textcolor{TensionOrange}{%
        \(\leftrightarrow\)\,\textbf{\scriptsize TENSION}%
    }%
    \enspace #1%
}

\newtcbox{\chip}[1][gray]{%
  on line,
  arc=1mm, 
  colback=#1!10!white, 
  colframe=#1!50!black, 
  before upper={\strut},
  boxrule=0.5pt, 
  boxsep=0pt,
  left=3pt, 
  right=3pt,
  top=0.25pt, 
  bottom=0.25pt
}

\usepackage{tcolorbox}

\usepackage{longtable}
\usepackage[table]{xcolor}
\usepackage{booktabs}
\usepackage{array}

\definecolor{sectionblue}{RGB}{224,236,249}
\definecolor{headingblue}{RGB}{31,99,154}

\begin{document}

\title[Design Patterns for Educational Technologies]{LINK: A Corpus-Grounded Design Pattern Library for Educational Technologies}


\author{Lorena Lyon}
 \orcid{0009-0004-4068-8062}
 \affiliation{%
  \institution{National Cheng Kung University}
   \country{Taiwan}
 }
 \email{pa6157075@gs.ncku.edu.tw}

\author{Hyoungwook Jin}
 \orcid{0000-0003-0253-560X}
 \affiliation{%
  \institution{University of Michigan}
   \country{USA}
 }
 \email{jinhw@umich.edu}

\author{Vincent Cavez}
 \orcid{0000-0002-6662-7362}
 \affiliation{%
  \institution{Stanford University}
   \country{USA}
 }
 \email{vcavez@stanford.edu}

\author{Roy Pea}
 \orcid{0000-0001-6301-3536}
 \affiliation{%
  \institution{Stanford University}
   \country{USA}
 }
 \email{roypea@stanford.edu}

\author{Xu Wang}
 \orcid{0000-0001-5551-0815}
 \affiliation{%
  \institution{University of Michigan}
   \country{USA}
 }
 \email{xwanghci@umich.edu}

\author{Hari Subramonyam}
 \orcid{0000-0002-3450-0447}
 \affiliation{%
  \institution{Stanford University}
   \country{USA}
 }
 \email{harihars@stanford.edu}

\renewcommand{\shortauthors}{Lyon, et al.}
\begin{abstract}
Designing learning interfaces is often an ill-structured problem: effective solutions depend on learning goals, subject matter, learner characteristics, and interaction context that are difficult to generalize. Designers must integrate knowledge from learning sciences, interaction design, and subject domains when translating learning goals into interface decisions. Yet, prior efforts to connect learning theory and design have largely focused on learning activities or specific technologies, leaving interface-level design knowledge fragmented across systems. In this work, we take a bottom-up approach to surface this knowledge from existing educational interfaces. We present \system (Linking INstructional Knowledge to interfaces), a library of 27 recurring interface patterns, organized under nine higher-level design commitments, which describe broader learning-science-grounded goals for interfaces. As part of developing \system, we conducted a user study (N=14) with designers to evaluate \system's generative power. Our findings suggest that \system helps designers generate interface ideas and articulate theory-grounded rationales for design decisions.

\end{abstract}


\begin{teaserfigure}
  \includegraphics[width=\textwidth]{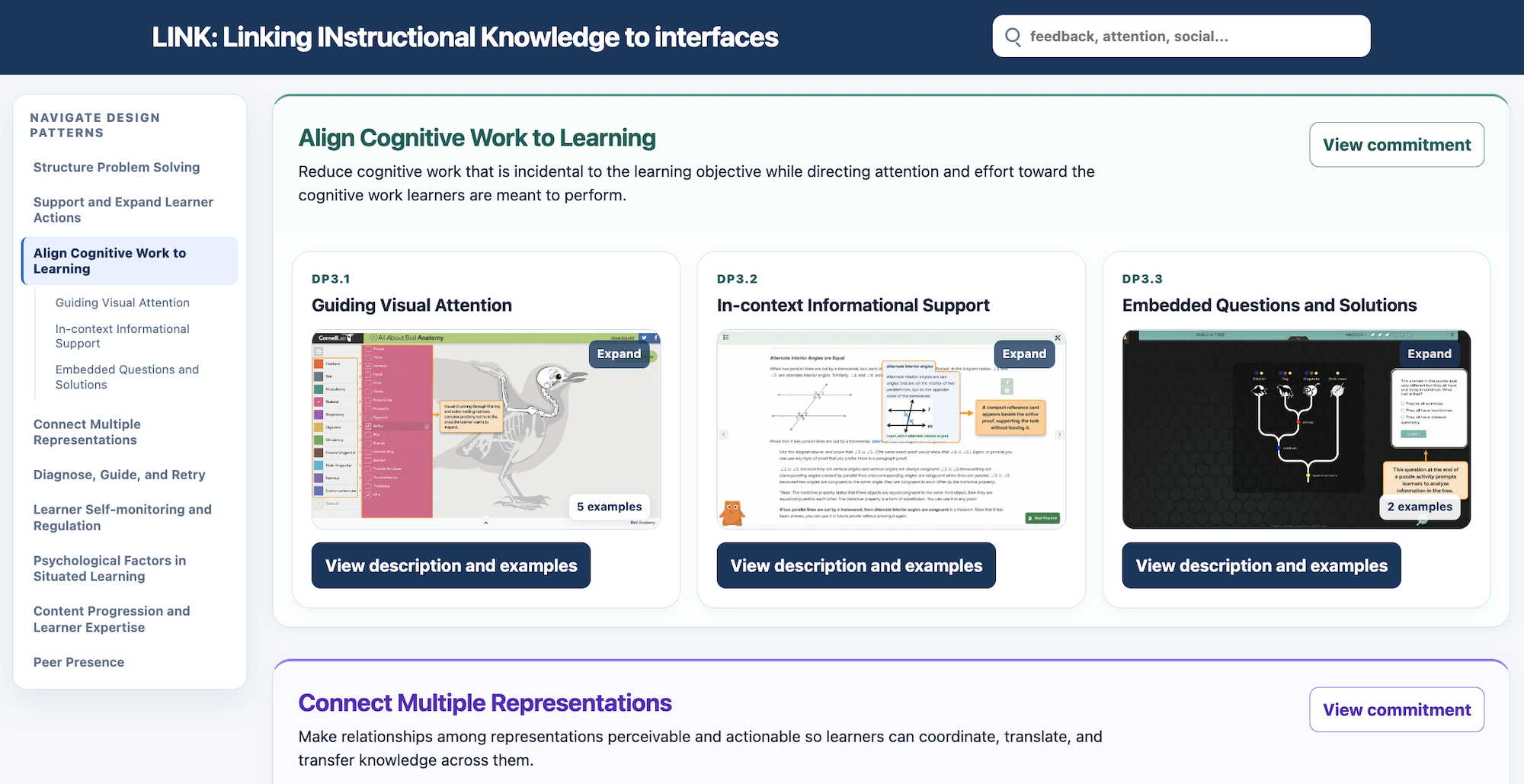}
   \caption{The \system web interface, which organizes 27 learning-interface design patterns under nine higher-level design commitments and provides examples, rationale, applicability, and limitations for each pattern.}
   \Description{Screenshot of a website showing a navigation bar on the left, and design patterns, which are organized into sections called "design commitments." Design patterns display a title, such as Decomposition of Steps, and a representative screenshot with annotations.}
   \label{fig:teaser}
 \end{teaserfigure}


\maketitle

\section{Introduction}

Digital learning tools such as interactive simulations, language-learning apps, math learning games, etc. \textit{translate} instructional goals into interfaces through which learners encounter content knowledge, take actions, and receive feedback. \textit{Designing} these learner-facing interfaces requires bringing together knowledge from the learning sciences, interaction design, and the subject domain of instruction. Consider a designer who creates an interface to help a novice learn fractions. Learning sciences may suggest that learners benefit from connecting symbolic expressions to visual representations and maintaining desirable difficulty when receiving support. Domain expertise is needed to ensure that those representations preserve the mathematical relationships the learner is meant to understand. The interaction design then determines how these ideas become concrete, i.e., whether learners manipulate pieces of a fraction model, enter a numerical answer, receive a hint after an error, or move between visual and symbolic representations. Each of these (seemingly small) interface decisions requires reasoning in learning science, domain, and design. More importantly, these design decisions are consequential and can shape learners' access, effort, persistence, and engagement. For example, in a large-scale study of an online mathematics and science platform, simply making the option to retry a failed problem more visually significant increased the students' likelihood of attempting the problem again~\cite{asher2026will}. 

For designers, translating learning goals into interface decisions can be challenging because there is often a many-to-many mapping between theories about how people learn and interface designs. These decisions often involve competing objectives. For example, providing guidance can help a novice begin a task while minimizing the opportunity for productive struggle. Simplifying an interface can reduce extraneous cognitive demands while also removing work central to the learning objective. On the other hand, increasing the choice of the learner can support agency while overwhelming learners who lack sufficient prior knowledge or self-regulation~\cite{bjork2011making,squires1999usability}. Learning-interface design is therefore an ill-structured design problem, with many of the characteristics associated with \textit{wicked design problems}~\cite{buchanan1992wicked}, i.e., there is rarely a single correct solution, design goals can conflict, and the appropriateness of a solution depends on the learner, learning objective, domain, and surrounding context. 

For such design problems, abstract principles and prescriptive guidelines are not enough. Designers face a multitude of plausible instructional and interaction choices~\cite{koedinger2013complexity}, while learning theories typically operate at a level of abstraction that leaves open how they should be implemented in concrete interfaces. Previous work has sought to make learning theory more actionable through mappings to learning activities and technological affordances, learning experience design, pedagogical pattern languages, and technology-specific design guidance~\cite{conole2004mapping,jeong2016collaboration,schmidthuang2022lxd,goodyear2005patterns,morwinters2007,tzortzoglou2025ar,mubarrat2026robotics,chowdhury2025disengagement}. However, these efforts operate largely above or within particular classes of technology, minimizing the focus on the recurring \textit{mechanisms} at the interface level through which learning principles become concrete in what learners see, do and receive in response. In practice, designers bridge this gap through combinations of intuition, precedent, prior experience, and subject-matter expertise \cite{zhou2023lxd, caskurlu2025data}. The resulting knowledge of how learning principles become interface decisions is often \textbf{embedded in design practice and existing products rather than explicitly articulated}.

Existing learning applications offer a rich source of such knowledge. Across products and domains, designers have developed concrete ways to handle key design issues for learning environments \cite{collins1996design} to structure problems, guide attention, coordinate representations, provide feedback, regulate learner agency, and support self-monitoring. In this work, we take a \textit{bottom-up approach} to examine existing interfaces to surface the design knowledge embedded in their recurring interface mechanisms. We curated 873 screenshots from 65 publicly available learner-facing applications spanning various subject domains and interaction formats. Following corpus-based approaches to design space analysis and pattern mining~\cite{grammel2013survey,satyanarayan2019reflections,gray2018dark}, we iteratively analyzed interface artifacts and identified 27 recurring interface patterns organized under nine higher-level \textit{design commitments}. We call this design knowledge \system (Linking INstructional Knowledge to interfaces). We use \textit{design commitments} to describe broader learning-oriented design goals that shape interface decisions and \textit{design patterns} to describe recurring interface mechanisms through which those commitments are realized. Importantly, we do not treat these patterns as universally effective or ``best'' designs. Instead, the library makes existing design choices legible by connecting interface mechanisms with the learning problems they address, their theoretical rationale, and the contexts and tensions surrounding their use.

Our study is guided by the following research questions:

\begin{enumerate}[label=\textbf{RQ\arabic*:}]
    \item What recurring interface patterns and higher-level design commitments characterize learner-facing digital learning applications?
    
    \item How do designers use and interpret a library of these patterns when reasoning about learning interface design?
\end{enumerate}

Based on the identified design commitments and patterns, we conducted a user study with 14 designers to examine how this intermediate design knowledge enters design practice. Designers used the pattern library to generate and extend interface ideas, reconsider emerging designs, and articulate rationales for their decisions. Their use of the library also surfaced challenges in making design knowledge actionable, including the importance of concrete examples, interpretable terminology, and ways to navigate a large collection of patterns. Our main contributions are: (1) \system: a corpus-grounded synthesis of 27 learning-interface design patterns organized under nine design commitments, and (2) an evaluation showing how designers use the pattern library to generate ideas, reason about decisions, and articulate design rationale.

\section{Related Work}

\subsection{Operationalizing Learning Theories}
The learning sciences offer a rich body of theory about how people learn. Even within computer science education, Szabo et al. identified 75 learning theories that span experiential, cognitive, and social perspectives~\cite{szabo2019learningtheories}. Learning theories in education have been said to be like toothbrushes: everyone needs one, but no one wants to use anyone else's. Previous work has translated broad theoretical traditions into instructional strategies~\cite{khalil2016} and linked types of knowledge to appropriate learning processes and instructional choices~\cite{koedinger2012kli}. However, we find that educational technologies are often weakly grounded in explicit theory, including virtual reality environments~\cite{marougkas2023vr} and educational games, whose pedagogical assumptions remain implicit in their mechanics~\cite{kacmaz2022games}. Previous work has sought to bridge theory and design. For example, Conole et al. map theoretical positions to e-learning activities~\cite{conole2004mapping}, while Jeong and Hmelo-Silver connect collaborative-learning needs to technological advantages and design strategies~\cite{jeong2016collaboration}. Other work provides theoretically-informed guidance for particular populations or technologies~\cite{vella2026attention,vendrell2026ai}. Learning experience design (LXD) integrates human-centered design with theoretically grounded learning goals~\cite{tawfik2022,schmidthuang2022lxd}. Central to LXD is the distinction between a learner and a generic user~\cite{soloway1994learner}: optimizing only for ease of use can remove challenge, structure, or cognitive activity important for learning~\cite{quintana2020lxd,squires1999usability}.

The remaining challenge for edtech UX designers is translating these principles into concrete interface decisions. Learning and instructional designers often rely on subject-matter experts, precedent, and professional judgment more than on codified theory or data~\cite{zhou2023lxd,caskurlu2025data}, while novices can struggle to expand learning goals into design moves~\cite{chang2020novice}. HCI has developed forms of intermediate design knowledge, including design principles~\cite{blairearly2008}, design patterns \cite{borchers2000pattern}, and generative and analytical frameworks~\cite{fang2026frameworks}. In educational design, Sandoval's design conjectures similarly connect the features designed with the hypothesized learning processes and outcomes~\cite{sandoval2004conjectures,sandoval2014conjecture}. We build on this tradition by using \textit{design commitments} to capture higher-level learning rationales that connect learning concerns to recurring interface mechanisms.

\subsection{Interface Design Knowledge for Learning Technologies}
HCI has long treated learners as a distinct user population~\cite{soloway1994learner}. Early work translated learning concerns into interface guidance, including metacognition-based guidelines~\cite{jones1995metacognitive}, usability criteria for e-learning systems~\cite{ardito2004usability}, and a scaffolding framework for inquiry software~\cite{quintana2004scaffolding}. Such work demonstrates the value of connecting learning activities and interface design, but typically remains anchored in a particular theory or class of system. More recent work shows that even small interface choices can be pedagogically consequential. A large-scale experiment found that a visual retry cue and persuasive trigger increased persistence after failure~\cite{asher2026persistence}; authoring systems encode pedagogical decisions into interfaces for educational simulations and games~\cite{kaputa2026simstep,yang2026language}; and interface presentation can shape learners' perceptions of the learning activity itself~\cite{tran2026scratch}. Design knowledge also emerges through situated practice, including multi-stakeholder co-design~\cite{mavri2025codesign} and negotiations between learning and UX designers~\cite{quintana2020lxd}. These approaches generate rich knowledge, but it often remains tied to the systems and settings in which it was produced.

A smaller body of research aggregates knowledge across systems. Studies have synthesized interaction designs in 62 mobile augmented-reality learning applications~\cite{tzortzoglou2025ar}, 95 studies of game-based robotics education~\cite{mubarrat2026robotics}, and 47 disengagement-support applications~\cite{chowdhury2025disengagement}, while other work has modeled Computer-Supported Collaborative Learning (CSCL) scripts \cite{dillenbourg2008mechanics}. These analyses demonstrate the value of recovering recurring design knowledge, but remain scoped to particular technologies or problem domains. We extend this approach across a wider range of learner-facing interfaces and examine how recurring mechanisms can be represented as reusable, learning-oriented design knowledge.

\subsection{Design Patterns as a Knowledge Representation}
Design patterns provide an established form for representing reusable knowledge in context-dependent design spaces. Originating in architecture~\cite{alexander1977language}, patterns later moved through software engineering~\cite{gamma1995patterns} into HCI~\cite{borchers2000patterns}. Unlike simple guidelines, patterns preserve a recurring problem, a possible solution, and the context, forces, and rationale surrounding its use~\cite{vanwelie2001patterns}; pattern catalogs have consequently become practical design resources~\cite{tidwell2005designing}. Patterns therefore narrow a design space without implying that one solution is universally correct. Education research has developed a parallel pattern tradition. The Pedagogical Patterns Project codified experienced teachers' practices~\cite{sharp2003pedagogical}; Goodyear positioned patterns between abstract theory and overly prescriptive guidance~\cite{goodyear2005patterns}; and Mor and Winters proposed them as a means of accumulating design knowledge in technology-enhanced learning~\cite{morwinters2007}. These patterns primarily describe pedagogical activities, social arrangements, and learning structures, whereas conventional HCI patterns describe interface solutions without their learning rationale or associated social context, such as the specification of participant roles. Interface-level patterns that preserve both the mechanism and its learning rationale remain comparatively underdeveloped as contributions to learning environment and UX design.

Corpus-based design analyses offer a way to recover such intermediate knowledge from existing artifacts. In data visualization, Grammel et al. synthesized approaches to visualization construction and their tradeoffs~\cite{grammel2013survey}, while Satyanarayan et al. compared visualization-authoring systems to expose recurring design choices~\cite{satyanarayan2019reflections}. However, observing a recurring pattern does not establish its effectiveness; instead, patterns make existing mechanisms, rationales, and tradeoffs available for dialogue and critical examination. Dearden and Finlay found limited evidence that patterns improve design outcomes and cautioned against adopting their form without their underlying values~\cite{dearden2006patterns}; Fang et al. similarly note that HCI often produces frameworks faster than it evaluates them~\cite{fang2026frameworks}. We respond by grounding patterns in recurring instances across independent systems and by examining how designers interpret and use the resulting library in an evaluation study.
\section{Method}

We used a corpus-based, bottom-up approach to surface recurring design knowledge embedded in existing learning interfaces. Our method combined qualitative design pattern mining with iterative, abductive analysis of interface artifacts. We began with observable interface mechanisms and learner interactions while drawing on learning-science concepts to interpret the learning problems and rationales which those mechanisms appeared to address. This approach is based on prior design-space analyses of existing systems~\cite{grammel2013survey,satyanarayan2019reflections,lee2024design}, pattern taxonomies derived from the corpus of deployed interfaces~\cite{gray2018dark}, and content analyses of educational applications~\cite{callaghan2018educational}. The analysis produced an initial set of patterns that we subsequently refined through independent reorganizing and an evaluation study with designers (Section~\ref{sec:validation}).

\subsection{Researcher Positionality}
Our analysis is shaped by the combined expertise of the research team in learning sciences, human-computer interaction, and learning technology design. Team members have worked in academic, nonprofit, and commercial educational-technology settings, and three have taught graduate-level courses in learning technology design that explicitly train students to translate learning theories into design artifacts. This combination of learning-science, HCI, and design expertise helped us interpret how interface mechanisms relate to learning goals and theoretical principles. At the same time, it may have made patterns aligned with theories and design practices familiar to us more readily visible. We treat the resulting patterns as interpretive constructions rather than objective features of the corpus, and we used independent regrouping and team discussion to surface and challenge these interpretations.

\begin{figure*}
    \centering
    \includegraphics[width=0.7\textwidth]{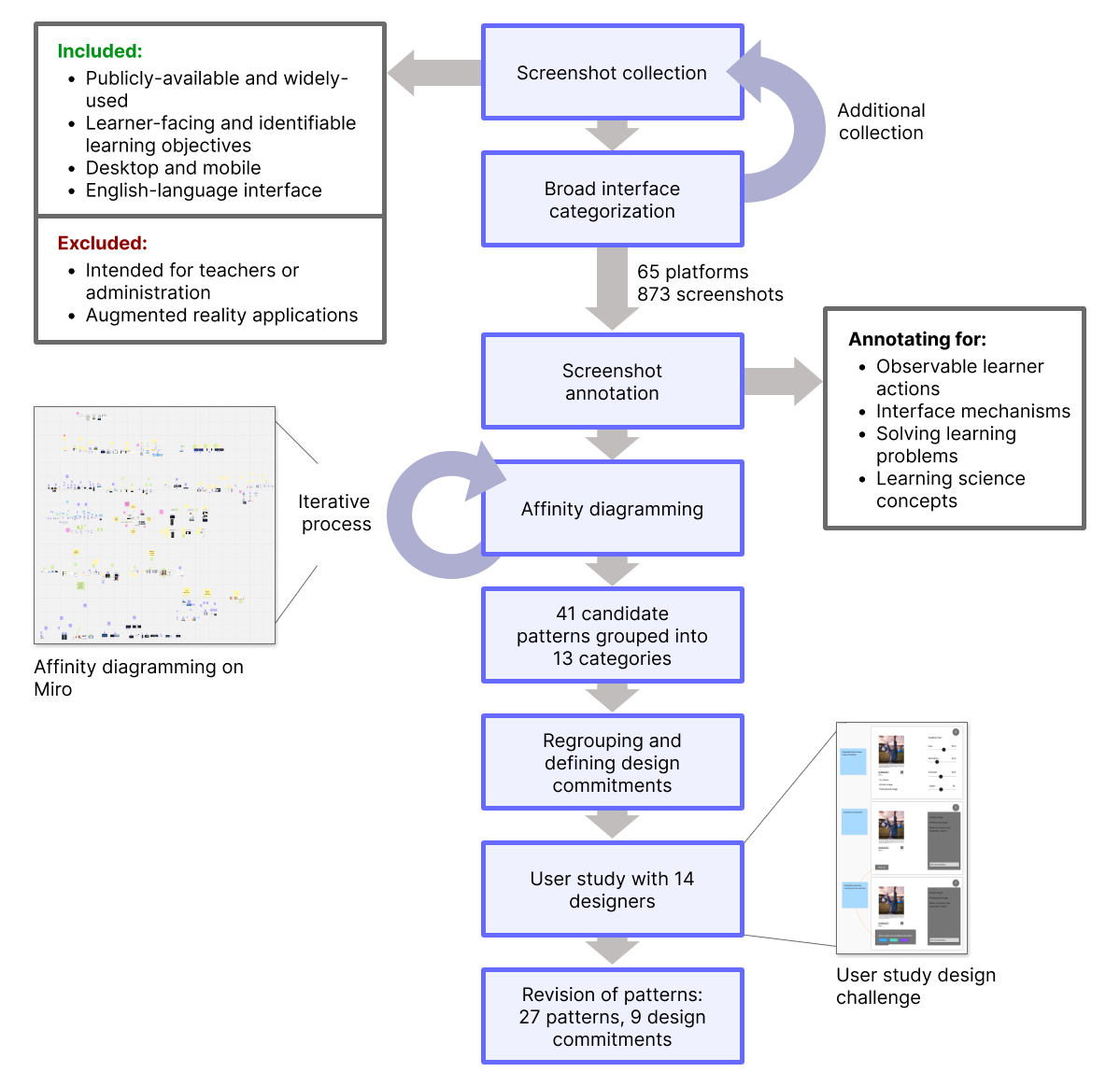}
    \caption{Overview of the process used to collect and analyze 873 screenshots from 65 platforms, develop preliminary design patterns, regroup and define design commitments, and refine design patterns through a study with 14 designers. The final collection included 27 patterns organized by nine design commitments.}
    \Description{Flowchart of the design-pattern development process: screenshots from 65 learner-facing platforms were collected using defined inclusion and exclusion criteria, broadly categorized, and annotated for learner actions, interface mechanisms, learning problems, and learning-science concepts. Iterative affinity diagramming of 873 screenshots produced 41 patterns in 13 categories, which were regrouped into design commitments, evaluated in a study with 14 designers, and revised into 27 patterns and nine design commitments.}
    \label{fig:Process}
\end{figure*}

\subsection{Corpus Collection}
Our goal is to capture a sufficiently broad range of learner-facing interfaces from which recurring design mechanisms can be identified. We use purposive sampling~\cite{palinkas2015purposeful,patton2014qualitative}, beginning with widely-used and publicly-available learning applications across desktop and mobile, including applications familiar to the authors and highly ranked applications in the Apple and Google app stores, an approach used in previous educational-application analyses~\cite{callaghan2018educational}. We subsequently broadened the corpus using Y Combinator's education-startup directory\footnote{\url{https://www.ycombinator.com/companies/industry/education/}}, the Accessible Digital Learning Portal\footnote{\url{https://accessibledigitallearning.org/catalog/}}, and other collections of learning technologies. As the analysis progressed, we intentionally added applications representing interface types and learning domains that were underrepresented in the existing corpus.

We \textbf{included} applications that (1) were primarily learner-facing and organized around identifiable learning objectives, (2) were publicly accessible through a mobile application or desktop browser, and (3) provided an English-language interface. We \textbf{excluded} systems primarily intended for teachers or administrators, such as classroom-management and teacher-authoring tools. We also excluded augmented-reality applications because this design space has recently been systematically characterized by Tzortzoglou et al.~\cite{tzortzoglou2025ar}.

Our primary analytic artifacts were interface screenshots~\cite{deka2017rico,dillman2018videogames}. Most screenshots were collected by researchers while walking through applications and their interactions, following the principles of the walkthrough method~\cite{light2018walkthrough}. When interfaces were inaccessible due to subscriptions or other access restrictions, we supplemented these materials with publicly available product previews, promotional screenshots, and examples from the Mobbin interface-reference site~\cite{mobbin}. We distinguish these sources because promotional and reference materials provide evidence of intended interface design rather than direct evidence of an application's complete interaction flow.

After an initial review and subsequent additions, the resulting corpus contained 873 screenshots from 65 \textit{platforms}. We define a platform as a cohesive digital learning experience organized around a distinct subject domain, learning objectives, content, tasks, and interaction patterns. For example, we treated Duolingo Math and Duolingo Music as separate platforms because they support different learning objectives through substantially different interfaces. The corpus spans conventional instructional platforms, learning games and simulations, programming environments, and AI-supported learning tools. It was designed for the breadth of interface mechanisms rather than statistical representativeness.

\subsection{Pattern Mining and Analysis}

\subsubsection{Initial Pattern Identification} 
We imported the corpus into a shared Miro workspace that offers a flexible canvas to spatially organize content and add notes and annotations on one integrated surface~\cite{miro}. We organized screenshots into broad interface formats, including conversational tutors, programming environments, structured-response activities, video-based lessons, simulations, and games, to assess the breadth of the corpus. Note that these categories were used only to identify coverage gaps and were not retained as analytic categories. When a category or interaction form appeared to be underrepresented, we collected additional applications and screenshots before continuing the analysis.

We then annotated screenshots for learner actions, interface mechanisms, the learning problem the mechanism appeared to address, and possible connections to learning-science concepts. Our unit of interpretation was not simply a visual component, but a relationship between a \textit{learner problem}, an \textit{interface mechanism}, and its plausible \textit{learning rationale}. For temporal mechanisms such as progressively revealing actions or changing representations across a learning sequence, we examined screenshots in the context of the walkthrough sequence rather than treating an isolated screen as sufficient evidence. Using affinity diagramming~\cite{scupin1997kj,lucero2015using}, we grouped annotations that reflected similar interface mechanisms and learning problems. We iteratively compared candidate groups between applications, split groups that captured meaningfully different mechanisms, and merged groups that represented variants of the same underlying design response. This process produced an initial set of 41 candidate patterns organized into 13 provisional groups.

\begin{figure*}
    \centering
    \includegraphics[width=\textwidth]{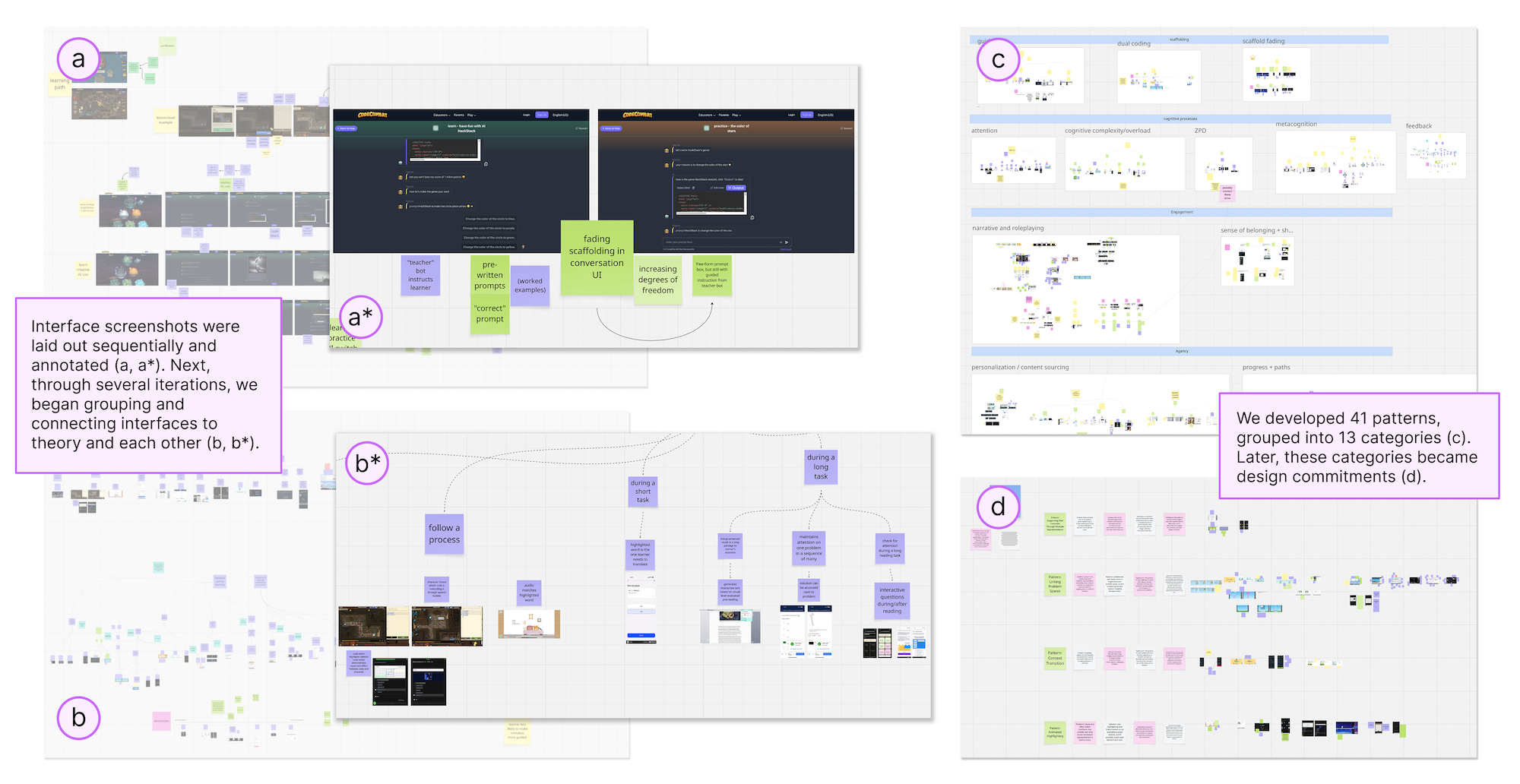}
    \caption{We annotated interface screenshots before using affinity diagramming to group interfaces and develop patterns. Later, we treated groupings as design commitments.}
    \Description{Four-stage overview of the design-pattern development process in Miro: interface screenshots were first arranged sequentially and annotated (a, a*), then iteratively grouped and connected to learning theory and one another (b, b*), producing 41 patterns organized into 13 categories (c), which were later consolidated into design commitments (d).}
    \label{fig:Miro-process}
\end{figure*}

\subsubsection{Developing Design Commitments}
Our initial organization grouped patterns under broad categories such as guidance and agency and, in some cases, arranged them along dimensions of implementation. Through continued analysis, we found that these dimensions often imposed a structure that was not supported consistently across patterns. We instead began treating the higher-level groupings as \textit{design commitments}: learning-oriented rationales that explain why different interface mechanisms may address related learning problems. To develop these commitments, three researchers independently regrouped the 41 candidate patterns according to the learning-oriented rationale each appeared to reflect. We then compared the independent groupings and discussed disagreements until reaching an interpretive consensus~\cite{mcdonald2019reliability}. Because the purpose of this stage was conceptual interpretation rather than assigning observations to a fixed coding scheme, we used consensus discussion rather than inter-rater reliability as the criterion for resolving differences.

For a candidate mechanism to be retained as a design pattern, we required evidence of recurrence across independent learning systems and sufficient information to characterize (1) the learner-facing problem, (2) the recurring interface mechanism, and (3) the learning rationale or tradeoff associated with the mechanism. In practice, we required instances on at least three independent platforms. This criterion distinguishes patterns from the idiosyncratic features of a single product. After developing the initial pattern library, we conducted a user study to gather feedback on the organization and presentation of the patterns. Following the study, we revisited overlapping and difficult-to-distinguish patterns in light of both the corpus and the interpretations of the participants. We merged conceptually redundant patterns and revised their organization and descriptions, resulting in the final set of 27 patterns organized under nine design commitments. Section~\ref{sec:validation} describes the study and the role it played in this refinement.
\section{Patterns for Learning Interface Design}

Our analysis produced 27 recurring interface patterns organized under nine \textit{design commitments}. For each commitment, we first synthesize the relevant learning-science perspective and the interface-level design problem that emerged across the corpus. We then summarize its constituent patterns in terms of three elements: the interface \textit{mechanism}, the conditions under which that mechanism may be useful, and the \textit{tension} or tradeoff introduced by its use. In each table, the mechanism is expressed as an interface-level design move followed by the change it produces in the learner's activity.

\subsection{Structure Problem Solving}
\label{sec:structure-problem-solving}

\designcommitment{\underline{DC1:} Expose enough of a task's structure to help learners make progress while preserving the problem-solving decisions they are meant to practice.}

Problem solving requires learners to represent a goal, identify pertinent information, select actions, and monitor their progress toward a solution~\cite{dewey1933,miller1960plans,polya1945how,simon1971human}. However, novices may have difficulty identifying productive approaches or structuring unfamiliar problems~\cite{bjorklund2013initial}. Step decomposition and worked examples can help students identify goals and
strategies for problem solving~\cite{catrambone1998subgoal, atkinson2000learning}. Cognitive apprenticeship similarly makes otherwise hidden expert processes visible by modeling and scaffolding learners through a process~\cite{collins1989cognitive}.

\begin{table*}
\caption{Patterns associated with the \textit{Structure Problem Solving}
design commitment.}
\label{tab:structure-problem-solving}
\small
\setlength{\tabcolsep}{8pt}
\renewcommand{\arraystretch}{1.30}

\begin{tabularx}{\textwidth}{
    >{\columncolor{PatternLight}}P{0.20\textwidth}
    P{0.37\textwidth}
    Y
}
\toprule
\rowcolor{CommitmentBlue}
\color{white}\textbf{Pattern} &
\color{white}\textbf{Mechanism: Design $\rightarrow$ Learning} &
\color{white}\textbf{Applicability \& Tension} \\
\midrule

\patternname{\underline{DP1.1} Decomposition of Steps}
&
\mechanism{Decomposes a problem into ordered subproblems}
{making intermediate goals and their sequence explicit to the learner.}
\corpusexample{AI tutoring systems decompose learner-supplied problems into
smaller problems and guide learners through them sequentially.}
&
\when{Learners may not yet know how to begin or structure a problem for which
an established solution process exists.}

\tension{Making the process visible can support progress, but may offload the
work of learning how to decompose problems independently. Learners may execute
individual steps correctly without understanding how the decomposition itself
was produced.}
\\

\addlinespace[4pt]

\patternname{\underline{DP1.2} Ideal Output Modeling}
&
\mechanism{Makes the target outcome visible while leaving the path open}
{orienting learners toward a goal without prescribing the intermediate
actions they should take.}
\corpusexample{Observed examples included an investing simulator displaying an
``optimal play-through'' trajectory, an RNA-folding game displaying target
structures, and a drawing interface placing a model artwork beneath the
learner's canvas.}
&
\when{Tasks are exploratory, puzzle-like, creative, or permit multiple valid
solution paths.}

\tension{A model can preserve productive freedom, but learners may still be
unable to identify how their current approach differs from the target or what
intermediate action would close the gap.}
\\

\addlinespace[4pt]

\patternname{\underline{DP1.3} Structured Representational Space}
&
\mechanism{Encodes domain relationships into the available interactions}
{allowing exploration while constraining actions that violate relevant
structural rules.}
\corpusexample{Mathematics interfaces encoded multiplication in manipulable
rows and columns, while block-based programming systems prevented learners
from connecting structurally incompatible blocks.}
&
\when{Important domain relationships can be meaningfully represented through
the objects and actions of the interface, including sandbox environments that
permit free construction within defined rules.}

\tension{Embedded constraints can make domain structure tangible, but they
may become opaque when learners cannot infer why an input or action is
unavailable.}
\\

\bottomrule
\end{tabularx}
\end{table*}

Across our corpus, interfaces translated these ideas by selectively exposing different parts of a problem's underlying structure. The three patterns differ in \textit{where that structure is externalized}. \textit{Decomposition of Steps} exposes the intermediate process; \textit{Ideal Output Modeling} exposes the desired endpoint while leaving the process open; and \textit{Structured Representational Space} embeds domain relationships into the actions that the interface permits. Thus, the commitment is to decide which aspects of the problem should be made explicit and which should remain for learners to infer, select, or construct themselves. Each choice leaves different reasoning for students to perform themselves. The design tension is between making enough structure visible to enable productive action and preserving enough uncertainty for learners to practice the intended problem-solving work.

\begin{figure*}
    \centering
    \includegraphics[width=\textwidth]{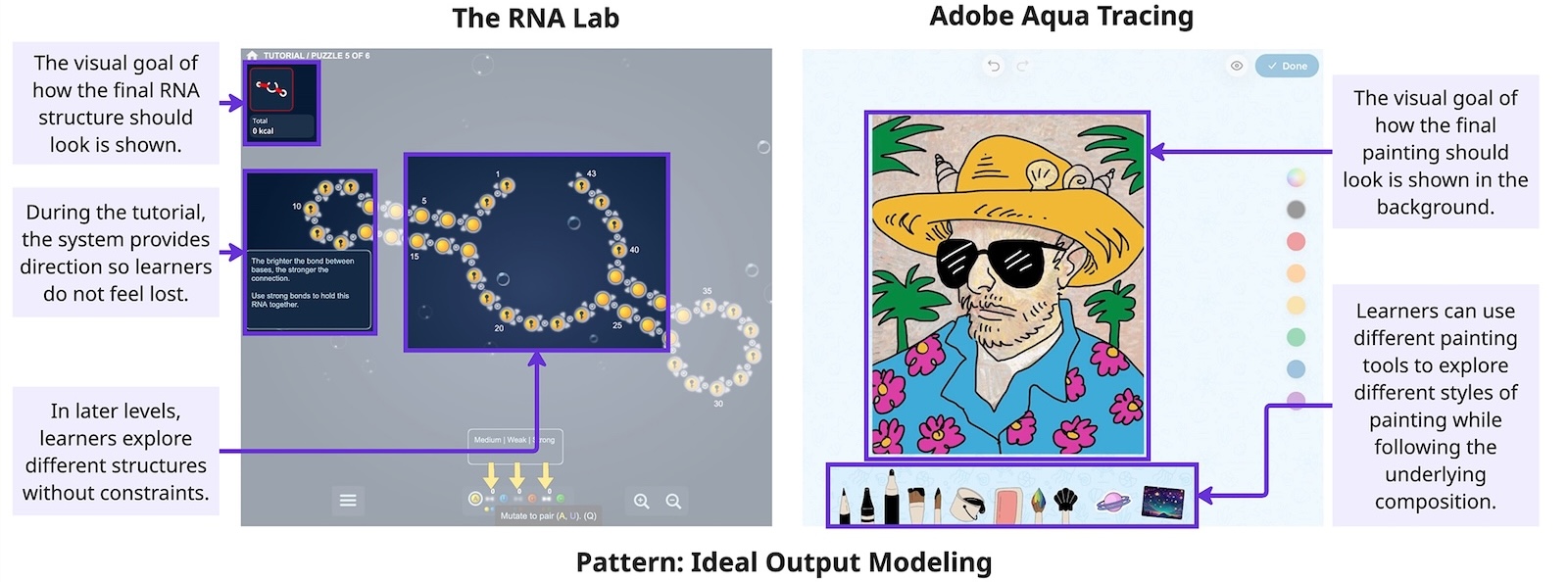}
    \caption{The RNA Lab and Adobe Aqua Tracing provide visible models of desired outcomes that orient learners while preserving flexibility in how they explore and produce their own solutions.}
    \Description{[DP1.2 Ideal Output Modeling] Screenshots from the RNA Lab and Adobe Aqua Tracing show visual models of desired outcomes. RNA Lab displays a small target RNA structure, an explanatory hint, and an editable molecular structure that learners can explore using different bond strengths. Adobe Aqua Tracing presents a completed tropical portrait as a visual guide and provides multiple drawing and painting tools. Both interfaces communicate an intended result while allowing learners flexibility in how they reach or reinterpret it.}
    \label{fig:ideal_output_modeling}
\end{figure*}

\begin{figure*}
    \centering
    \includegraphics[width=\textwidth]{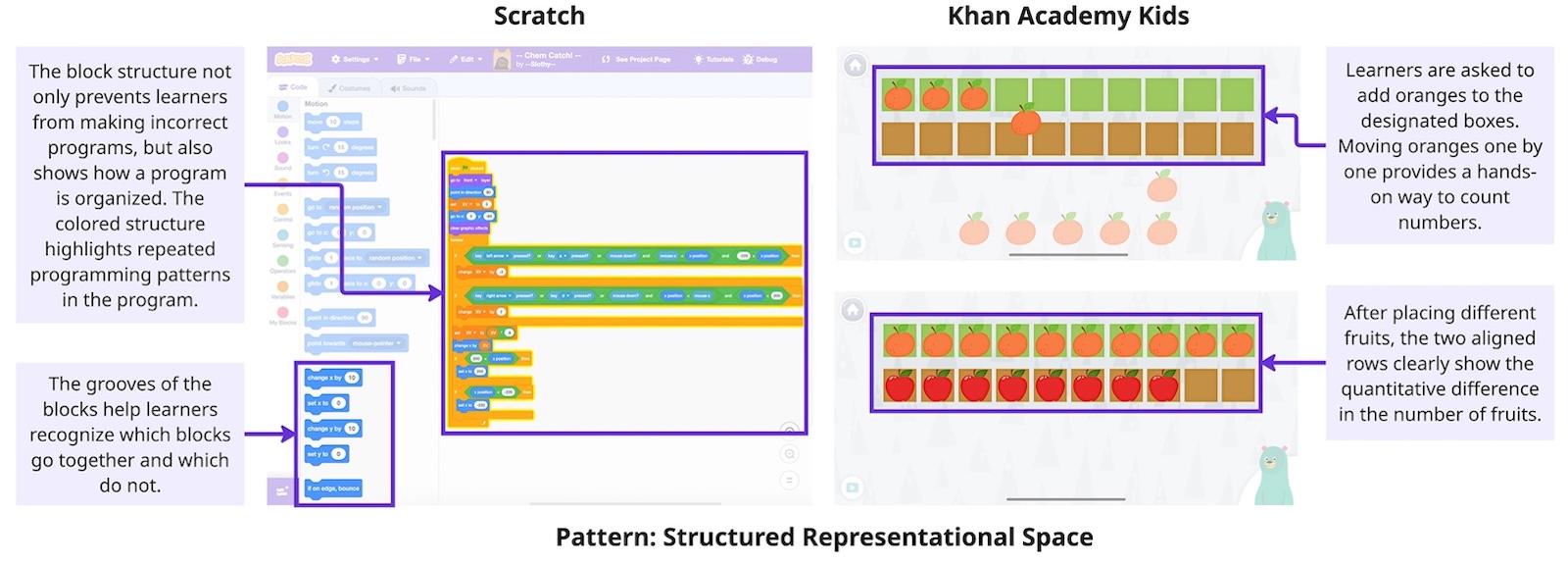}
    \caption{Scratch and Khan Academy Kids organize learning activities within constrained visual spaces that make underlying structures and relationships explicit while reducing opportunities for invalid actions.}
    \Description{[DP1.3 Structured Representational Space] Screenshots from Scratch and Khan Academy Kids illustrate structured visual spaces for learning. Scratch presents color-coded programming blocks whose shapes indicate which commands can connect; an assembled program makes nested and repeated patterns visible while helping prevent invalid combinations. Khan Academy Kids uses aligned boxes into which learners drag oranges and apples. The two rows support hands-on counting and make differences between quantities visually apparent.}
    \label{fig:structured_representational_space}
\end{figure*}

\subsection{Support and Expand Learner Actions}
\label{sec:support-expand-actions}

\designcommitment{\underline{DC2:} Calibrate the learner's action space to their current
competence by initially guiding or constraining action and progressively increasing autonomy as support becomes less necessary.}

Learning interfaces often present learners with an \textit{action space}: the set of inputs through which they can act on content. Open or complex action spaces may require subject-matter knowledge, interface familiarity, and self-regulation that learners have not yet developed. A learning scaffold is a temporary support that can be removed once the learner has gained sufficient mastery~\cite{wood1976scaffolding,pea2004social}. Scaffolding can enable learners to perform beyond their unaided competence by allowing the system to perform a part of the task~\cite{van2010scaffolding}, but this support should ultimately contribute to independent performance rather than becoming a permanent crutch~\cite{pea2004social}.

Throughout the corpus, this scaffolding logic was translated directly into the actions that learners could take. \textit{Suggested Actions} preserve the broader action space, but make a productive subset of possible actions more salient. \textit{Unlocking Actions}, in contrast, change the boundaries of the action space itself by initially withholding some forms of interaction and making them available later. The patterns differ in whether the interface changes \textit{which actions appear preferable} or \textit{which actions are possible}. In both cases, support is intended to be temporary. The design question is when and how responsibility for navigating the interaction should shift from the interface toward the learner.

\begin{table*}
\caption{Patterns associated with the \textit{Support and Expand Learner
Actions} design commitment.}
\label{tab:support-expand-actions}
\small
\setlength{\tabcolsep}{8pt}
\renewcommand{\arraystretch}{1.30}

\begin{tabularx}{\textwidth}{
    >{\columncolor{PatternLight}}P{0.20\textwidth}
    P{0.37\textwidth}
    Y
}
\toprule
\rowcolor{CommitmentBlue}
\color{white}\textbf{Pattern} &
\color{white}\textbf{Mechanism: Design $\rightarrow$ Learning} &
\color{white}\textbf{Applicability \& Tension} \\
\midrule

\patternname{\underline{DP2.1} Suggested Actions}
&
\mechanism{Surfaces productive actions within an otherwise available action space}
{reducing the need to determine how to begin while preserving alternative
actions.}
\corpusexample{Evolution Lab and Beastbox suggest initial actions;
Codecademy provides pre-filled planning tasks; AI tutoring systems suggest
possible questions or intentions for learners to pursue.}
&
\when{Learners are entering an unfamiliar or open-ended activity and may not
yet recognize productive ways to proceed.}

\tension{Suggestions can orient learners while preserving choice, but
persistent or overly specific suggestions can replace rather than support
learner decision-making.}
\\

\addlinespace[4pt]

\patternname{\underline{DP2.2} Unlocking Actions}
&
\mechanism{Restricts available actions initially and expands them over time}
{matching the complexity of the interaction space to the learner's developing
competence.}
\corpusexample{Duolingo Music begins with only a few piano keys
and progressively exposes additional keys and notation; CodeCombat HackStack, a chat-based learning
system, moves from predefined message chips to free-form input.}
&
\when{The complete interaction space would initially overwhelm novice learners
or be difficult to use productively, especially in simulations, sandboxes,
chat, or agent interfaces.}

\tension{Expanding freedom too early can overwhelm novices, whereas expanding
it too late can frustrate learners whose knowledge or interface competence has
already surpassed the available controls.}
\\

\bottomrule
\end{tabularx}
\end{table*}

\begin{figure*} [t]
    \centering
    \includegraphics[width=\textwidth]{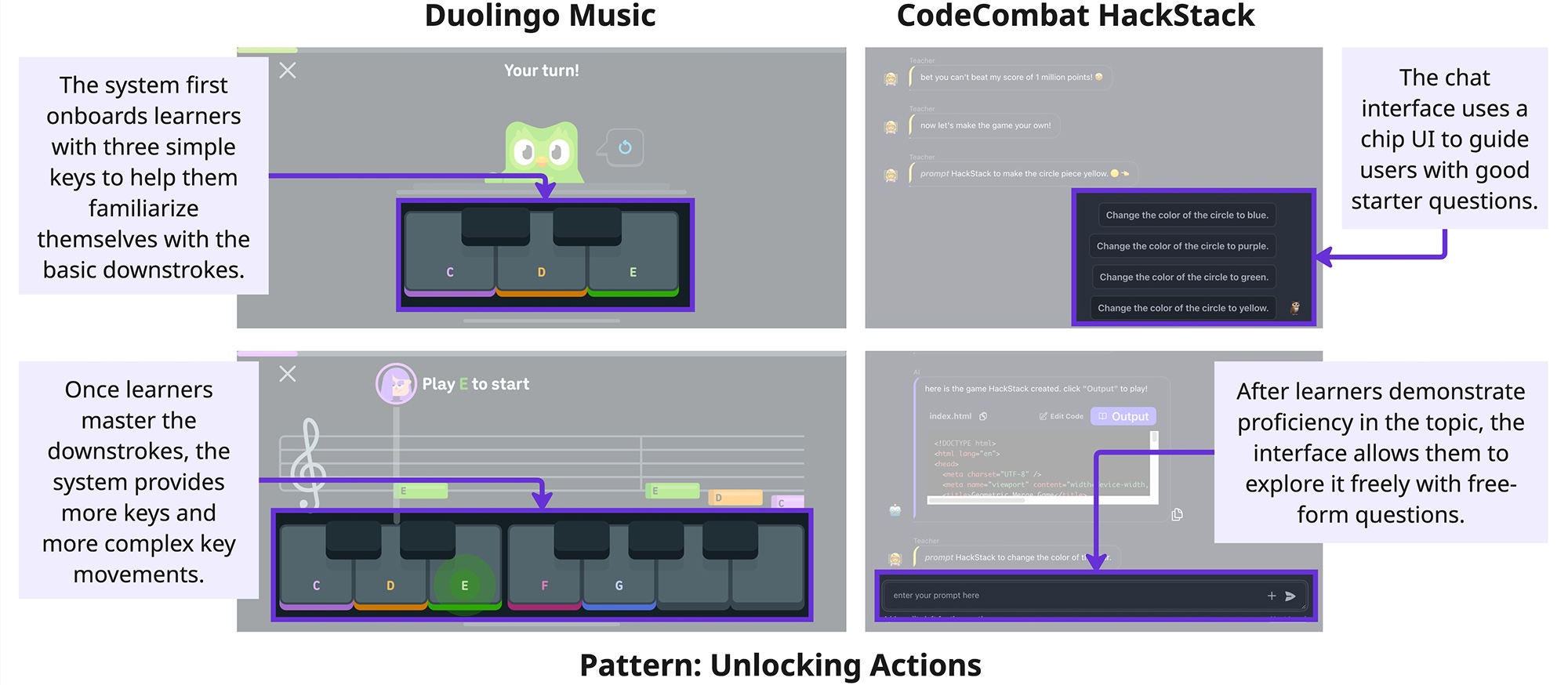}
    \caption{Duolingo Music and CodeCombat HackStack progressively expand the actions available to learners, beginning with guided choices and unlocking more complex or open-ended interactions as proficiency develops.}
    \Description{[DP2.2 Unlocking Actions] Two top-to-bottom sequences demonstrate progressive access to learning actions. Duolingo Music initially limits learners to three piano keys for practicing basic downstrokes, then expands the keyboard and introduces more complex note movements. CodeCombat HackStack initially provides selectable prompt suggestions for modifying a game, then offers an unrestricted text field for free-form prompts. In both examples, the interface expands learners' options as they gain familiarity or proficiency.}
    \label{fig:unlocking_actions}
\end{figure*}

\subsection{Align Cognitive Work to Learning}
\label{sec:align-cognitive-work}

\designcommitment{ \underline{DC3:} Reduce cognitive work that is incidental to the learning objective while directing attention and effort toward the cognitive work learners are meant to perform.}

During learning, learners engage in cognitive work such as processing information, directing attention, searching, integrating, and reasoning. Because working memory and attention are limited resources, demands introduced by an interface can compete with those required by the learning task. Cognitive load theory distinguishes the demands of processing learning materials from the extraneous load created by their presentation
~\cite{schnotz2007cognitiveload,sweller2020cognitive}. Mayer and Moreno's Spatial Contiguity Principle similarly shows that keeping related instructional elements physically close reduces unnecessary visual search and integration~\cite{mayer2003nine}. Sustained tasks can also benefit from interventions that redirect attention when it starts to drift~\cite{thomson2015resource}.

The current literature framing also connects this commitment to adjunct questions and interspersed testing. Rothkopf (1970) described the ``Mathemagenic Effect of Selective Attention,'' in which questions embedded within a reading passage function as attention guides and shift learners from passive exposure toward active search~\cite{rothkopf1970concept}. Likewise, this work draws a connection to ``progressive disclosure,''
attributed to Van Dam (1988), in which interface capabilities are gradually revealed so that users do not need to process the full complexity of a system at once~\cite{van1988hypertext}. 

Since learning itself requires cognitive work, these patterns distinguish
work that advances the learning objective from work introduced incidentally by the interface or by prerequisite information. Interfaces can offload some of these processing demands into external representations
~\cite{schnotz2007cognitiveload,clark1998extended,clark2008supersizing}, but cognitive load should not be minimized indiscriminately. \textit{Guiding Visual Attention} reduces incidental search; \textit{In-context Informational Support} externalizes secondary knowledge at the point of need; and \textit{Embedded Questions and Solutions} place prompts or support within the content context where learners can use them without
reconstructing what they were previously attending to.

\begin{table*}[H]
\caption{Patterns associated with the \textit{Align Cognitive Work to
Learning} design commitment.}
\label{tab:align-cognitive-work}
\small
\setlength{\tabcolsep}{8pt}
\renewcommand{\arraystretch}{1.30}

\begin{tabularx}{\textwidth}{
    >{\columncolor{PatternLight}}P{0.20\textwidth}
    P{0.37\textwidth}
    Y
}
\toprule
\rowcolor{CommitmentBlue}
\color{white}\textbf{Pattern} &
\color{white}\textbf{Mechanism: Design $\rightarrow$ Learning} &
\color{white}\textbf{Applicability \& Tension} \\
\midrule

\patternname{\underline{DP3.1} Guiding Visual Attention}
&
\mechanism{Organizes, separates, or highlights interface information}
{reducing incidental visual search and directing attention toward
task-relevant features.}
\corpusexample{Observed mechanisms included color-coded anatomical
categories, spatially organized matching tasks, separate ``build'' and
``learn'' tabs for AI-assisted coding, and highlighting of key terms during
an explanation.}
&
\when{Locating or filtering interface elements is not itself part of the
learning objective, or when several modes or large sets of options must be
made navigable.}

\tension{Reducing search can preserve attention for learning, but may remove
useful discrimination work, impose misleading categories, hide relevant
context across views, or dilute salience when too many elements are
highlighted.}
\\

\addlinespace[4pt]

\patternname{\underline{DP3.2} In-context Informational Support}
&
\mechanism{Places optional secondary information at the point of need}
{offloading terminology, syntax, formulas, or calculations that would
otherwise block the focal task.}
\corpusexample{Examples include hover descriptions of programming function
arguments, automatic calculation of secondary financial quantities,
sentence-simplification controls, and illustrated definitions of unfamiliar
geometry terms.}
&
\when{Peripheral knowledge is necessary for participation but is not central
to what learners are intended to practice, particularly for mixed-level
audiences.}

\tension{Whether a demand is ``extraneous'' depends on the learning goal.
Providing vocabulary may appropriately support a speaking simulation, while
providing complete example phrases may offload too much of the target task.}
\\

\addlinespace[4pt]

\patternname{\underline{DP3.3} Embedded Questions and Solutions}
&
\mechanism{Interleaves questions or nearby solutions with extended content}
{redirecting attention while the relevant reading, video, problem, or context
remains salient.}
\corpusexample{Reading systems insert comprehension questions during and
after passages; an evolutionary-tree game asks learners to interpret
information after puzzles; repeated problem sequences often place solutions
directly after the corresponding problem.}
&
\when{Learners engage with extended readings, videos, problem sets, or other
activities in which attention may drift over time.}

\tension{Well-placed questions can restore attention and encourage active
processing, whereas poorly timed questions or solutions can interrupt flow or
prematurely disclose support.}
\\

\bottomrule
\end{tabularx}
\end{table*}

\begin{figure*} [t]
    \centering
    \includegraphics[width=\textwidth]{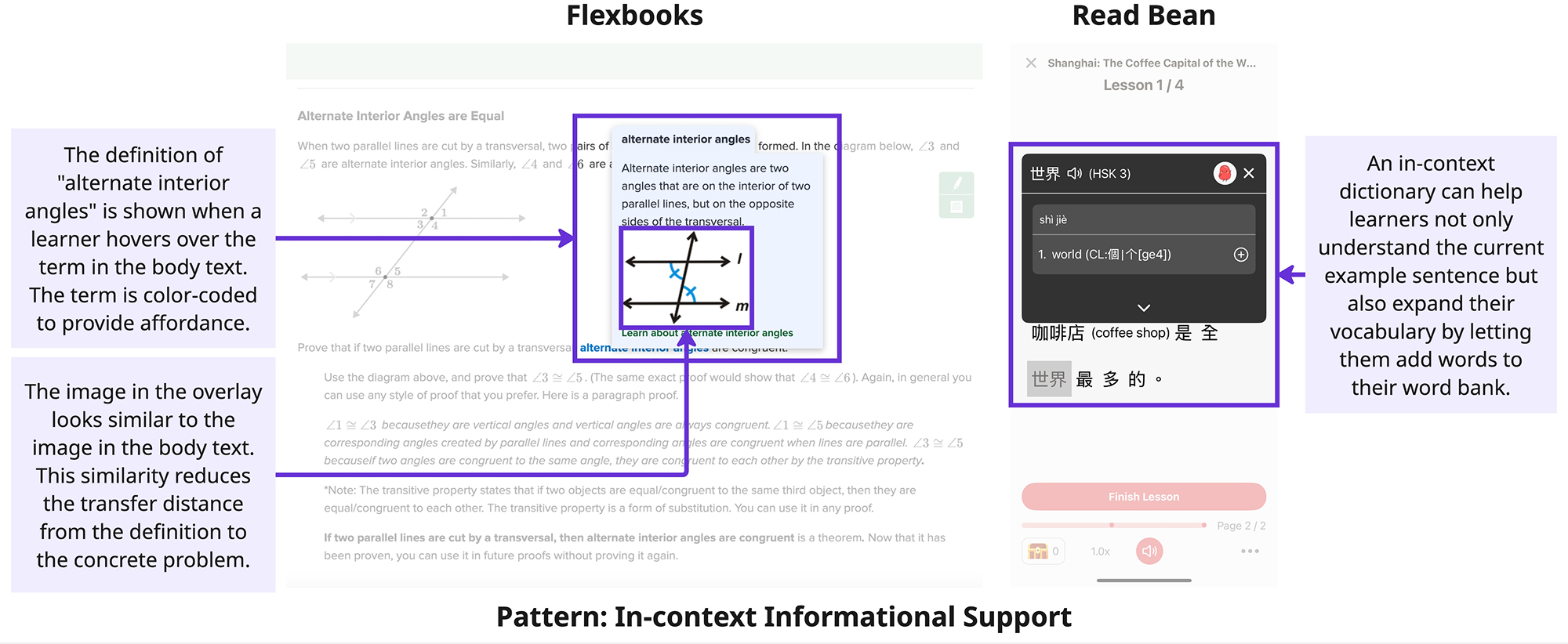}
    \caption{Flexbooks and Read Bean provide definitions and explanatory resources within the learning activity, helping learners resolve knowledge gaps without leaving the immediate task context.}
    \Description{Two examples illustrate informational support embedded directly within a learning activity. In Flexbooks, hovering over the color-coded term "alternate interior angles" opens a definition accompanied by a diagram resembling the geometry problem in the main text. This visual correspondence helps learners connect the concept to its application. In Read Bean, selecting the Chinese word opens an in-context dictionary showing its pronunciation, meaning, and an option to add it to a personal word bank, supporting both sentence comprehension and vocabulary development.}
    \label{fig:placeholder}
\end{figure*}

Across the three patterns, interfaces take responsibility for different forms of incidental cognitive work: \textit{finding} relevant information, \textit{retrieving} secondary information, or \textit{maintaining} attention through an extended activity. However, the same interface move can be productive or counterproductive depending on the learning objective. An aid is useful when it removes work peripheral to that objective; it can undermine learning when performing that work is itself what the learner is intended to practice.

\subsection{Connect Multiple Representations}
\label{sec:multiple-representations}

\designcommitment{\underline{DC4} Make relationships among representations perceivable and actionable so learners can coordinate, translate, and transfer knowledge across them.}

\begin{table*}
\caption{Patterns associated with the \textit{Connect Multiple
Representations} design commitment.}
\label{tab:multiple-representations}
\small
\setlength{\tabcolsep}{8pt}
\renewcommand{\arraystretch}{1.30}

\begin{tabularx}{\textwidth}{
    >{\columncolor{PatternLight}}P{0.22\textwidth}
    P{0.35\textwidth}
    Y
}
\toprule
\rowcolor{CommitmentBlue}
\color{white}\textbf{Pattern} &
\color{white}\textbf{Mechanism: Design $\rightarrow$ Learning} &
\color{white}\textbf{Applicability \& Tension} \\
\midrule

\patternname{\underline{DP4.1} Supporting New Concepts Through Multiple Representations}
&
\mechanism{Pairs verbal information with a corresponding visual}
{providing more than one representational route through which a new concept
can be encoded or recalled.}
\corpusexample{Clusive, a reading system, adds illustrative icons to selected text; an
early-learning interface pairs a letter and sound with familiar images.}
&
\when{New concepts or vocabulary have a direct and meaningful visual
correspondence and memorization or initial encoding is central to the task.}

\tension{Complementary representations can enrich encoding, while decorative,
loosely related, or redundant imagery can introduce distraction or additional
cognitive load.}
\\

\addlinespace[4pt]

\patternname{\underline{DP4.2} Linking Problem Spaces}
&
\mechanism{Requires mapping between representations within the same task}
{making coordination between visual, concrete, and symbolic forms part of
the learner's problem solving.}
\corpusexample{Mathematics tasks link illustrated quantities to numbers,
require learners to select symbolic expressions for sets of objects, or use
manipulable objects to solve division problems.}
&
\when{The learning objective involves moving between concrete, visual, or
symbolic representations, particularly in spatial and quantitative tasks.}

\tension{Requiring mapping can promote integration, but learners may still
struggle to identify the correspondence between pictorial and mathematical
representations~\cite{ainsworth2006mers}. A second representation may add
little value when it merely duplicates information.}
\\

\addlinespace[4pt]

\patternname{\underline{DP4.3} Highlighting Processes with Multiple Representations}
&
\mechanism{Synchronizes corresponding elements as a process unfolds}
{making temporal and causal relationships across representations easier to
trace.}
\corpusexample{Scratch, Brilliant, and CodeCombat highlight the lines of code
responsible for corresponding steps in a visualized process.}
&
\when{Dynamic representations change together over time and have a stable
correspondence that can be meaningfully highlighted.}

\tension{Synchronization can reveal causal relationships, but rapid,
ambiguous, or many-to-many mappings may be difficult to follow. Learners may
need controls to pause, replay, or step through the process.}
\\

\addlinespace[4pt]

\patternname{\underline{DP4.4} Context Transition}
&
\mechanism{Removes or changes a supporting representation while preserving
the underlying problem structure}
{moving learners toward more abstract or independently usable forms of the
same knowledge.}
\corpusexample{Duolingo Math's ratio problems progress from image--number matching to
numerical calculation; another mathematics interface moves from measurements
taken from a picture into a geometry problem using those measurements.}
&
\when{The learning objective is transfer from a concrete or multiple
representation toward a symbolic, abstract, or single-channel representation.}

\tension{A transition supports transfer only if learners recognize the
underlying isomorphism; otherwise the new representation may be experienced
as a separate and unrelated problem.}
\\

\bottomrule
\end{tabularx}
\end{table*}

Multiple representations, such as text, imagery, symbolic notation, and interactive visualizations, can provide learners with different ways of accessing and storing knowledge. Dual Coding Theory and Cognitive Theory of Multimedia Learning motivate combining verbal and visual representations~\cite{paivio1991dual, mayer2022cognitive}. Multiple representations can also provide complementary information or support different forms of problem solving~\cite{ainsworth2006mers}. However, learners do not necessarily integrate knowledge across representations merely because they are presented together~\cite{ainsworth2006mers}.

Across our corpus, interfaces established integral relationships among representations. \textit{Supporting New Concepts Through Multiple Representations} pairs representations when a concept is introduced; \textit{Linking Problem Spaces} requires learners to map between representations as part of completing a task; \textit{Highlighting Processes with Multiple Representations} synchronizes corresponding representations over time; and \textit{Context Transition} gradually removes or changes a representation while preserving the underlying problem structure. Together, these patterns make representational coordination part of the learner's activity rather than assuming that integration will occur automatically.

The four patterns describe different forms of representational connection: \textit{pairing}, \textit{mapping}, \textit{synchronizing}, and \textit{transitioning}. Here, the design commitment is more than simply ``provide multiple representations.'' It concerns how the interface makes relationships between representations available for learners to perceive, act on, and eventually carry into contexts where one of those representations
is absent.

\begin{figure*}
    \centering
    \includegraphics[width=\textwidth]{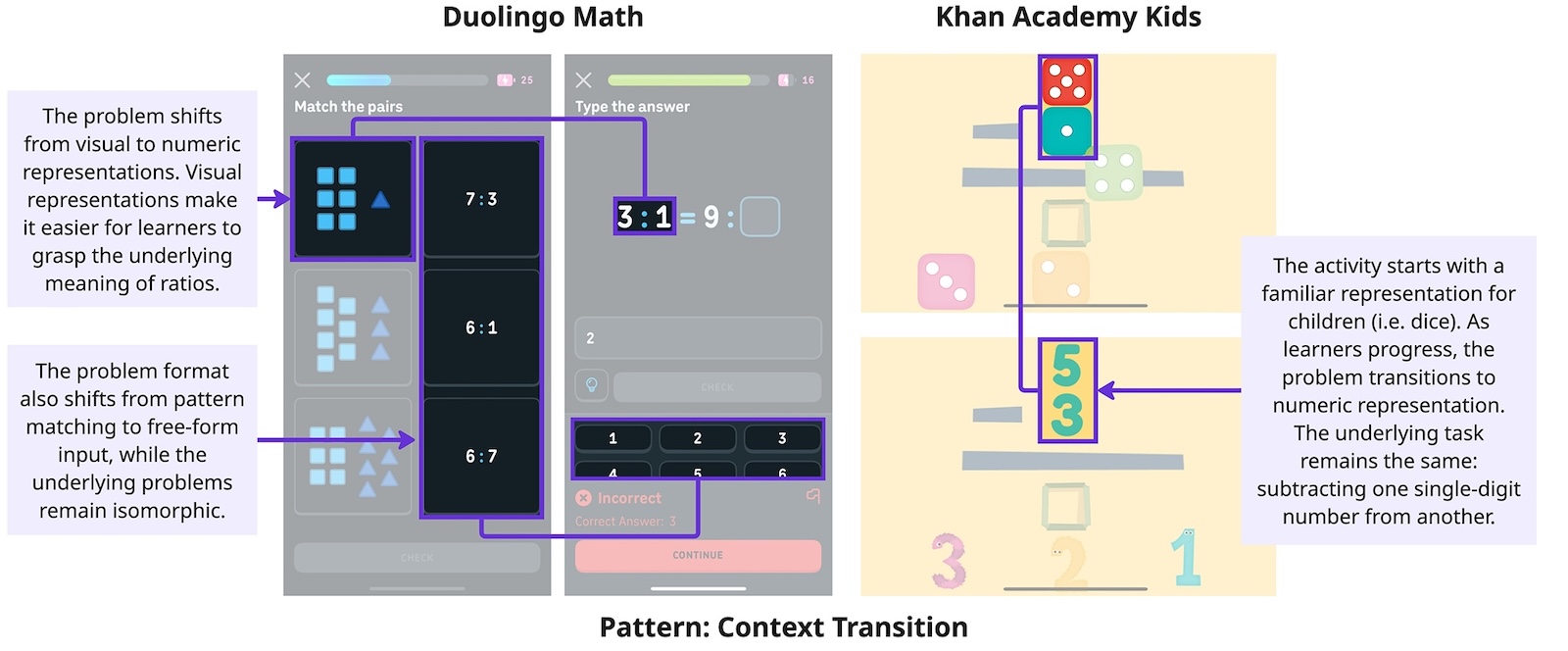}
    \caption{Duolingo Math and Khan Academy Kids gradually transition learners from familiar visual representations to more abstract numerical forms and from recognition-based tasks to open-ended responses.}
    \Description{[DP4.4 Context Transition] Two examples show learning activities transitioning from familiar or concrete representations to more abstract ones. Duolingo Math first represents ratios with groups of shapes in a matching task, then expresses the same relationship numerically in a free-response equation. Khan Academy Kids similarly moves from subtraction represented with dice to subtraction represented with Arabic numerals. In both cases, the underlying problem structure remains similar while the representation or response format becomes more abstract.}
    \label{fig:context_transition}
\end{figure*}

\subsection{Diagnose, Guide, and Retry}
\label{sec:diagnose-guide-retry}

\designcommitment{\underline{DC5:} Calibrate the specificity of feedback to what learners need to recognize an error, revise their reasoning, and make productive use of remaining attempts.}

In learning science, corrective feedback is important because unaddressed errors can continue to interfere with subsequent learning~\cite{fu2022effects}. This work further motivates the feedback as a way to prevent an incorrect response from continuing to compete with the representation of the intended knowledge~\cite{butler2008correcting}. Feedback can also help learners recognize the gap between their current performance and a learning objective~\cite{sadler1989formative, hattie2007power}. Effective feedback should provide information that learners can use to revise their understanding or strategy~\cite{garciajimenez2015feedback}.

Throughout the corpus, responding to an error was not a single design choice. The interfaces varied both \textit{how much diagnostic information they revealed} and \textit{what the learners could do next}. \textit{Binary Feedback} communicates correctness with minimal explanation; \textit{Explanatory Feedback} makes corrective information explicit; \textit{Iterative Feedback} withholds a full solution so that learners can diagnose and retry; and \textit{Generative Explanations} adapt the explanation to an individual learner's current
question or difficulty.

\begin{table*}
\caption{Patterns associated with the \textit{Diagnose, Guide, and Retry}
design commitment.}
\label{tab:diagnose-guide-retry}
\small
\setlength{\tabcolsep}{8pt}
\renewcommand{\arraystretch}{1.30}

\begin{tabularx}{\textwidth}{
    >{\columncolor{PatternLight}}P{0.20\textwidth}
    P{0.37\textwidth}
    Y
}
\toprule
\rowcolor{CommitmentBlue}
\color{white}\textbf{Pattern} &
\color{white}\textbf{Mechanism: Design $\rightarrow$ Learning} &
\color{white}\textbf{Applicability \& Tension} \\
\midrule

\patternname{\underline{DP5.1} Binary Feedback}
&
\mechanism{Signals correctness immediately and unambiguously}
{allowing learners to distinguish a correct response from an error before
moving on.}
\corpusexample{Observed in simple, closed-ended questions through symbols such
as check marks and crosses and through conventional correctness colors.}
&
\when{Correctness can be evaluated clearly, and the task primarily assesses
retrieval or other closed-ended responses.}

\tension{Binary feedback closes the correctness loop efficiently but conveys
little information about why an answer was wrong, making it insufficient when
errors arise from misconceptions rather than lapses in recall.}
\\

\addlinespace[4pt]

\patternname{\underline{DP5.2} Explanatory Feedback}
&
\mechanism{Explains the source of an error or demonstrates correct application}
{providing information learners can use to revise a misconception or
strategy.}
\corpusexample{Programming challenges use explanatory code review to suggest
improvements or alternative approaches.}
&
\when{Knowing only that a response is incorrect is insufficient because the
task depends on conceptual understanding or a complex procedure.}

\tension{More explanation can clarify misconceptions, but its appropriate
depth depends on task complexity, and it may reveal reasoning learners could
otherwise perform themselves.}
\\

\addlinespace[4pt]

\patternname{\underline{DP5.3} Iterative Feedback}
&
\mechanism{Reveals partial diagnostic information without disclosing the full solution}
{preserving opportunities for learners to diagnose, revise, and try again.}
\corpusexample{An interactive logic puzzle indicates which rules were followed
correctly without explicitly identifying the complete correction.}
&
\when{Tasks permit multiple attempts or allow the system to identify partially
correct aspects of a learner response.}

\tension{Partial feedback can preserve productive retry, but learners may
instead guess repeatedly rather than integrate and reason from the feedback.}
\\

\addlinespace[4pt]

\patternname{\underline{DP5.4} Generative Explanations}
&
\mechanism{Generates explanations responsive to the learner's current problem}
{adapting feedback to different questions, misconceptions, or levels of
understanding.}
\corpusexample{Brilliant's AI learning assistant can respond to questions
about a coordinate plane while highlighting the same axes or points with
which the learner is interacting.}
&
\when{Content is complex, and different learners may require substantially
different explanations or forms of support.}

\tension{Adaptive explanation can increase relevance, but generated content
may be inaccurate or pedagogically misaligned. Continuously available AI
assistance may also reduce productive struggle or create undesirable
dependencies.}
\\

\bottomrule
\end{tabularx}
\end{table*}

Viewed together, these patterns form an information gradient rather than a simple progression from ``less'' to ``more'' feedback. More information is not always more useful when learners still have productive work they can do themselves. The interface must coordinate the specificity of feedback with the next action available to the learner: move on, revise, retry, or seek a more individualized explanation.

\begin{figure*}
    \centering
    \includegraphics[width=\textwidth]{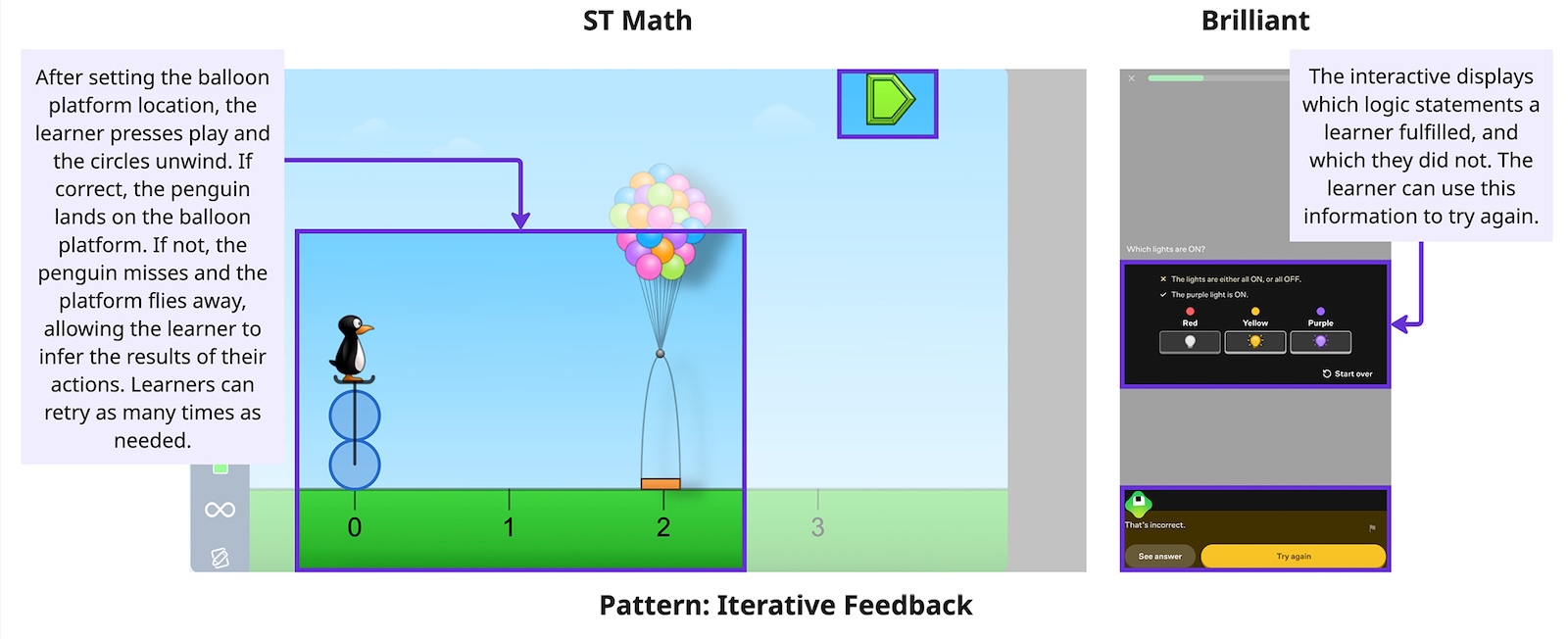}
    \caption{ST Math and Brilliant provide iterative feedback that helps learners evaluate their responses and refine them across multiple attempts.}
    \Description{[DP5.3 Iterative Feedback] Annotated screenshots illustrate iterative feedback in ST Math and Brilliant. In ST Math, the learner positions a balloon platform and presses play, causing a set of circles beneath a penguin to unwind. The penguin either lands on the platform or misses it, visually communicating the outcome and allowing retry. In Brilliant, the interface marks which conditions in a logic problem the learner did or did not satisfy, identifies the response as incorrect, and offers an option to try again.}
    \label{fig:iterative-feedback}
\end{figure*}

\subsection{Learner Self-monitoring and Regulation}
\label{sec:self-monitoring}

\designcommitment{\underline{DC6:} Externalize moments and information that help learners assess what they know, reflect on ongoing understanding, and monitor progress across time.}

Metacognition refers to the self-regulating cognitive processes through which learners monitor and manage their own learning ~\cite{flavell1979metacognition,brown1987metacognition,schraw1995metacognitive}. Metacognitive strategies can help learners self-evaluate and direct their learning efforts more effectively ~\cite{bjork2013self}. This work additionally notes that learners may mistake ease or fluency~\cite{roediger2006test} of processing~\cite{jemstedt2018ease} for genuine understanding when explicit opportunities for reflection are absent. Azevedo and Hadwin argue that computer-based environments can provide explicit scaffolds for learner monitoring and control
~\cite{azevedo2005metacognition}.

Across our corpus, interfaces created such opportunities at different timescales. \textit{Pre-Task Knowledge Assessment} makes a learner's starting state an object of inspection; \textit{Within-Task Knowledge Reflection}
introduces moments for monitoring during or around an activity; and \textit{Post-Task Learning Journey Overviews} aggregate activity and performance across longer periods. The common mechanism is not that the system performs
self-regulation for the learner, but that it makes information or reflective moments more explicit so that learners can regulate their own activity.

\begin{table*}
\caption{Patterns associated with the \textit{Learner Self-monitoring and
Regulation} design commitment.}
\label{tab:self-monitoring}
\small
\setlength{\tabcolsep}{8pt}
\renewcommand{\arraystretch}{1.30}

\begin{tabularx}{\textwidth}{
    >{\columncolor{PatternLight}}P{0.22\textwidth}
    P{0.35\textwidth}
    Y
}
\toprule
\rowcolor{CommitmentBlue}
\color{white}\textbf{Pattern} &
\color{white}\textbf{Mechanism: Design $\rightarrow$ Learning} &
\color{white}\textbf{Applicability \& Tension} \\
\midrule

\patternname{\underline{DP6.1} Pre-Task Knowledge Assessment}
&
\mechanism{Elicits a diagnostic or self-assessment before activity}
{making the learner's perceived or demonstrated starting state explicit.}
\corpusexample{Interfaces use content quizzes, Likert-style questionnaires
about attitudes or preconceptions, or brief questions about familiarity with
a topic.}
&
\when{Tasks depend on prerequisite knowledge, learners enter with different
levels of experience, or an initial estimate can meaningfully shape later
content or support.}

\tension{Assessment can promote awareness and calibration, but self-assessment
may be inaccurate, brief diagnostics may provide only a partial estimate, and
additional assessment can feel burdensome if its purpose is unclear.}
\\

\addlinespace[4pt]

\patternname{\underline{DP6.2} Within-Task Knowledge Reflection}
&
\mechanism{Prompts learners to articulate, categorize, or inspect their understanding}
{turning reflection on current knowledge into an explicit part of the
activity.}
\corpusexample{Examples include guided reflection questions after reading,
sorting vocabulary or flashcards into knowledge categories, and templates that
help learners organize their thoughts.}
&
\when{Tasks are complex or extended and learners benefit from monitoring what
they understand, do not understand, or have learned.}

\tension{Structured prompts can create opportunities for reflection, but
generic or poorly integrated prompts can feel superficial or add friction to
the learning activity.}
\\

\addlinespace[4pt]

\patternname{\underline{DP6.3} Post-Task Learning Journey Overviews}
&
\mechanism{Aggregates activity and performance across time}
{making longer-term patterns of participation or progress visible to the
learner.}
\corpusexample{Progress dashboards and activity metrics summarize learner
activity across sessions.}
&
\when{Learners return to a domain across multiple sessions or work toward
long-term goals such as an examination or a specific skill set.}

\tension{Visible metrics can support monitoring, but activity measures may
poorly represent actual understanding and can shift learner attention toward
maximizing the metrics the interface happens to expose.}
\\

\bottomrule
\end{tabularx}
\end{table*}

The three patterns place metacognitive support \textit{before}, \textit{within}, and \textit{across} learning episodes. This temporal distinction changes what learners are being invited to inspect, i.e., their starting knowledge, their current understanding, or their trajectory over time. In each case, the interface turns otherwise internal or diffuse information into something that can be explicitly considered by the learner.

\begin{figure*}
    \centering
    \includegraphics[width=\textwidth]{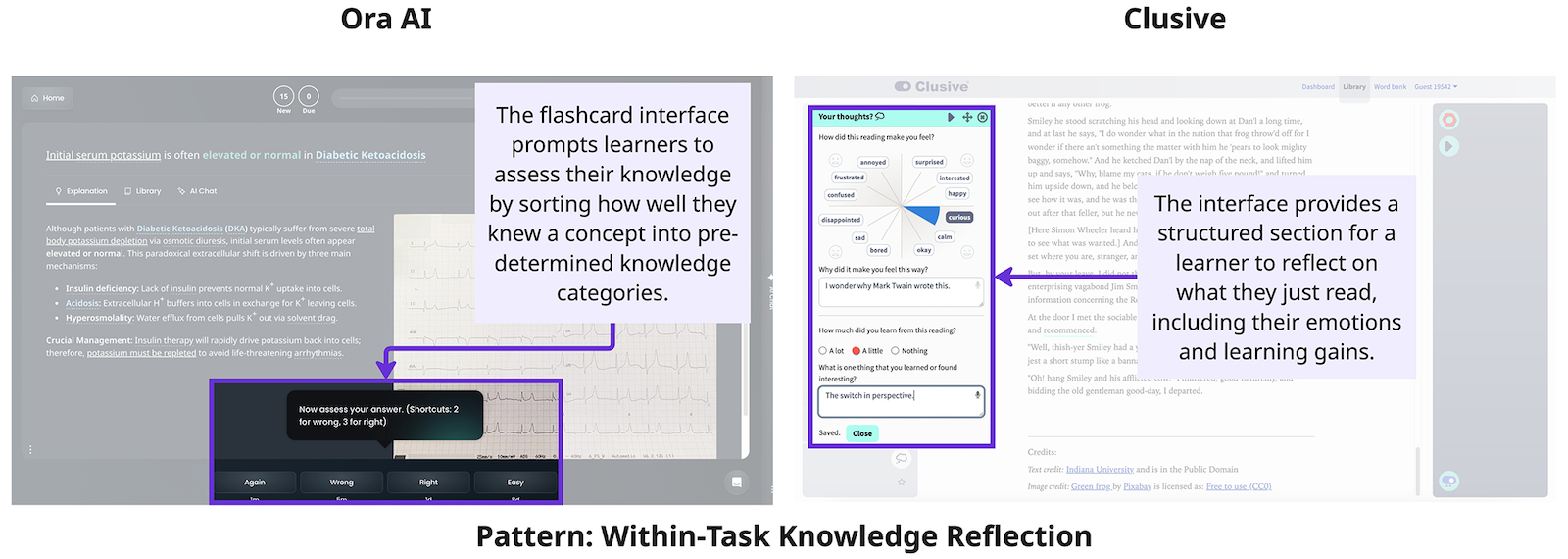}
    \caption{Ora AI and Clusive both structure opportunities for learners to reflect on their knowledge during a task.}
    \Description{[DP6.2 Within-Task Knowledge Reflection] Two examples showing how interfaces structure within-task knowledge reflection. In a flashcard explanation, Ora AI prompts learners to classify how well they knew the flashcard concept using four pre-determined knowledge categories: Again, Wrong, Right, and Easy. After a reading task, Clusive provides a structured panel for learners to report their emotions, learning gains, questions, and takeaways.}
    \label{fig:within-task-knowledge-reflection}
\end{figure*}

\subsection{Psychological Factors in Situated Learning}
\label{sec:psychological-situated}

\designcommitment{\underline{DC7:} Modulate psychological distance and role involvement so situated practice can balance emotional safety, meaningful consequence, and personal investment.}

Situated learning theory contends that knowledge is situated within the contexts and activities in which it is developed and used~\cite{brown1989situated}. Many systems in our corpus constructed such contexts through storytelling and role-playing. We theorize that this design space also intersects with construal-level theory, which describes how psychological distance---including social, spatial, and temporal distance---changes how situations are construed~\cite{trope2010construal}. Digital contexts can further provide relatively low-risk opportunities for learners to act unsuccessfully and revise their behavior, creating conditions that may support productive failure~\cite{kapur2008productive}.

The two patterns position learners differently relative to the simulated situation. \textit{Creating Psychological Distance} places the learner outside or one step removed from the focal consequences, for instance by asking them to observe a character or decide on another's behalf. \textit{Role-play} instead places the learner within a role and increases direct involvement in the scenario. Both use contextual framing to make practice possible, but one primarily creates \textit{distance from} the situation while the other creates
\textit{identification with} it.

\begin{table*}
\caption{Patterns associated with the \textit{Psychological Factors in
Situated Learning} design commitment.}
\label{tab:psychological-situated}
\small
\setlength{\tabcolsep}{8pt}
\renewcommand{\arraystretch}{1.30}

\begin{tabularx}{\textwidth}{
    >{\columncolor{PatternLight}}P{0.21\textwidth}
    P{0.36\textwidth}
    Y
}
\toprule
\rowcolor{CommitmentBlue}
\color{white}\textbf{Pattern} &
\color{white}\textbf{Mechanism: Design $\rightarrow$ Learning} &
\color{white}\textbf{Applicability \& Tension} \\
\midrule

\patternname{\underline{DP7.1} Creating Psychological Distance}
&
\mechanism{Positions the learner as an observer or proxy decision-maker}
{reducing the personal immediacy or perceived consequence of a sensitive
situation or failure.}
\corpusexample{A story-based system for neurodivergent children presents
another character's experiences; a finance game asks learners to make
decisions for a virtual pet; another social-learning activity asks learners
to make decisions for fictional characters.}
&
\when{Learning involves socially, emotionally, behaviorally, or financially
sensitive situations where direct first-person framing may feel threatening,
judgmental, or personally consequential.}

\tension{Distance can make difficult material safer to engage with, but can
also reduce personal relevance, investment, or accountability and make
transfer to the learner's own life more difficult.}
\\

\addlinespace[4pt]

\patternname{\underline{DP7.2} Role-play}
&
\mechanism{Places the learner inside a simulated role or situation}
{increasing personal involvement while allowing rehearsal in a comparatively
low-risk environment.}
\corpusexample{Read Bean and Codecademy support simulated conversations such
as language practice and job interviews; other applications place learners
in first-person science roles investigating astronomy, climate, or
conservation.}
&
\when{Skills benefit from situated practice, but authentic rehearsal is
high-stakes, inaccessible, psychologically challenging, or difficult to
arrange.}

\tension{Role-play can make practice approachable and motivating, but
simulations may simplify or misrepresent real situations and therefore
reinforce misconceptions or underprepare learners for real-world conditions.}
\\

\bottomrule
\end{tabularx}
\end{table*}

These patterns illustrate that contextualization does not imply a single preferred degree of immersion. A situation can support learning because it creates distance from the learner, or because it creates a stronger sense of participation. The relevant tension is between \textit{psychological safety and distance} on one side and \textit{identification, consequence, and personal relevance} on the other.

\begin{figure*}
    \centering
    \includegraphics[width=\textwidth]{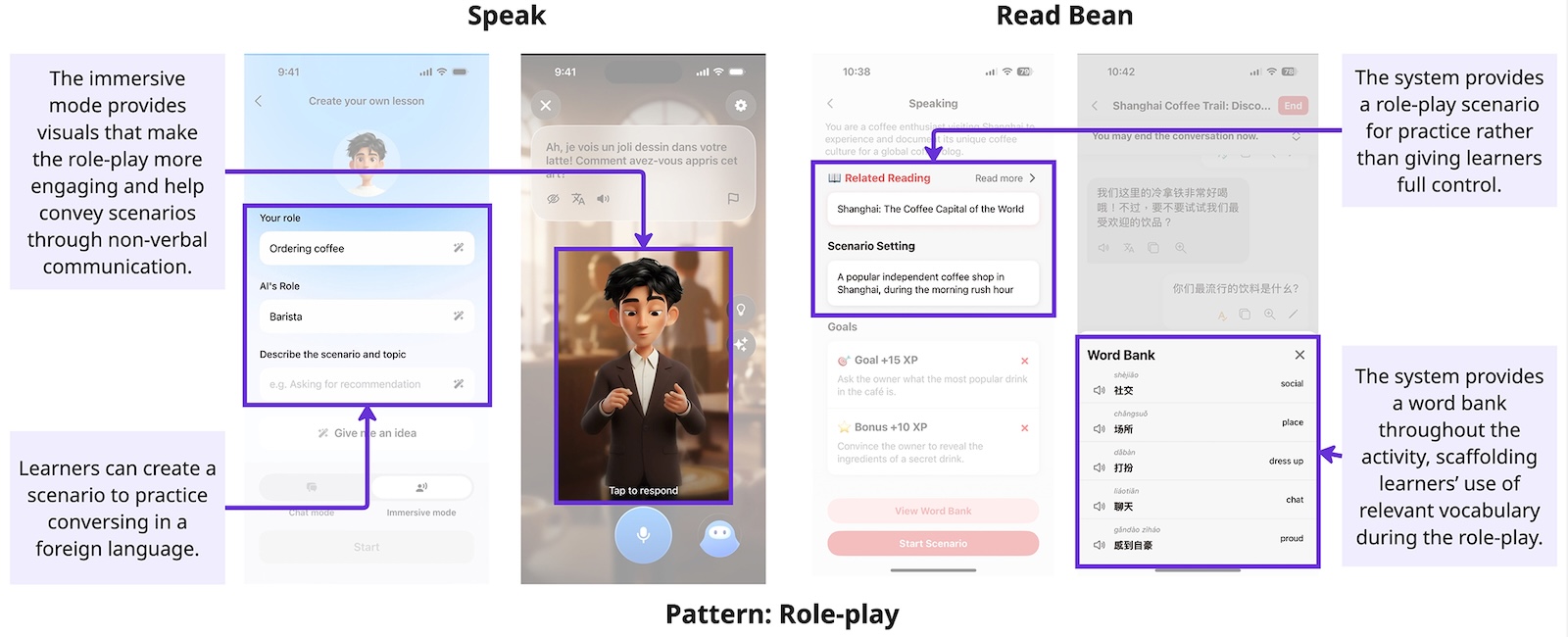}
    \caption{Speak and Read Bean support contextualized language practice through configurable role-play scenarios, immersive representations, explicit goals, and in-activity linguistic scaffolds.}
    \Description{[DP7.2 Role-play] Two sets of mobile screenshots illustrate scaffolded language role-play. In Speak, learners define their own role, the AI's role, and a scenario, then practice with an animated character in an immersive setting. In Read Bean, learners enter a predefined coffee-shop scenario supported by related reading and explicit goals. A word bank remains available during the conversation, providing vocabulary that learners can use in their responses.}
    \label{fig:roleplay}
\end{figure*}

\subsection{Content Progression and Learner Expertise}
\label{sec:content-progression}

\designcommitment{\underline{DC8:} Distribute control over content selection, sequencing, and
difficulty in relation to learners' domain knowledge and capacity for
self-direction.}

Choosing how to progress through content requires learners to set goals, identify relevant information, monitor their understanding, and adjust their strategies~\cite{winne1998studying}. Learners with limited domain knowledge
may lack the discernment needed to determine what content is relevant, how the materials relate, or what prerequisites they have not yet mastered. Learners with less-developed metacognitive skills may similarly struggle to evaluate their understanding and select an effective learning path~\cite{bjork2013self}. As learners develop knowledge and self-regulation, they may be better positioned to make decisions about content selection and
sequence~\cite{kalyuga2003expertise}.

Across the corpus, interfaces distributed responsibility for those decisions in different ways. \textit{Defined Learning Paths and Content} place sequencing primarily with the system. \textit{Customizable Paths and Content}
give the learner greater control over what content enters the learning experience and what direction it takes. \textit{Adaptive Learning Paths} retain control over the system while using information about the learner to alter content or difficulty. These patterns differ not simply in how ``personalized'' they are, but in \textit{who makes progression decisions} and what information grounds those decisions.

\begin{table*}
\caption{Patterns associated with the \textit{Content Progression and Learner
Expertise} design commitment.}
\label{tab:content-progression}
\small
\setlength{\tabcolsep}{8pt}
\renewcommand{\arraystretch}{1.30}

\begin{tabularx}{\textwidth}{
    >{\columncolor{PatternLight}}P{0.22\textwidth}
    P{0.35\textwidth}
    Y
}
\toprule
\rowcolor{CommitmentBlue}
\color{white}\textbf{Pattern} &
\color{white}\textbf{Mechanism: Design $\rightarrow$ Learning} &
\color{white}\textbf{Applicability \& Tension} \\
\midrule

\patternname{\underline{DP8.1} Defined Learning Paths and Content}
&
\mechanism{Pre-structures content into a visible sequence}
{externalizing decisions about prerequisites, ordering, and what the learner
should encounter next.}
\corpusexample{Khan Academy and Duolingo organize material into linear
visualized paths that signpost upcoming concepts and activity types such as
videos, practice, or challenges.}
&
\when{Learners are new to a domain or the material has meaningful curricular,
prerequisite, or established instructional sequences.}

\tension{Defined paths reduce the burden of sequencing, but may constrain
learners who already possess relevant knowledge, encounter material that is
too easy, or have goals that do not match the predetermined path.}
\\

\addlinespace[4pt]

\patternname{\underline{DP8.2} Customizable Paths and Content}
&
\mechanism{Lets learners select, supply, or generate the content they pursue}
{giving learners greater control over the goals, materials, and contexts that
shape the learning experience.}
\corpusexample{Clusive lets learners upload their own reading materials;
CapWords builds vocabulary from learner-captured images; Studdy and Brainly
generate explanations, guides, and practice from learner questions or
uploaded study materials.}
&
\when{Learners can make meaningful choices about their goals, possess
specialized questions or immediate learning needs, or bring personally
relevant materials that conventional authoring cannot anticipate.}

\tension{Greater agency can improve relevance, but learners with limited
domain knowledge or self-regulation may provide insufficient direction.
AI-generated paths may also contain inaccuracies, omit prerequisites, or
present content at an inappropriate level.}
\\

\addlinespace[4pt]

\patternname{\underline{DP8.3} Adaptive Learning Paths}
&
\mechanism{Uses information about the learner to adjust content or difficulty}
{retaining system structure while tailoring what the learner encounters to an
estimated level of expertise.}
\corpusexample{Read Bean uses a vocabulary assessment to estimate learner
level, assign appropriately leveled reading tasks, and scaffold vocabulary
above that estimated level.}
&
\when{Learners enter with heterogeneous levels of expertise and performance,
or diagnostic information can provide a meaningful basis for adjusting
difficulty.}

\tension{Adaptation can improve fit while preserving structure, but inaccurate
learner models can result in incorrectly leveled materials or content that is
poorly aligned with learner goals.}
\\

\bottomrule
\end{tabularx}
\end{table*}

The three patterns distribute control differently between learner and system: \textit{system-defined}, \textit{learner-directed}, or \textit{system-adaptive}. None resolves the problem of progression in every context. Instead, each makes a different assumption about whether the learner, a predefined curriculum, or information inferred by the system is best positioned to determine what should come next.

\begin{figure*}
    \centering
    \includegraphics[width=\textwidth]{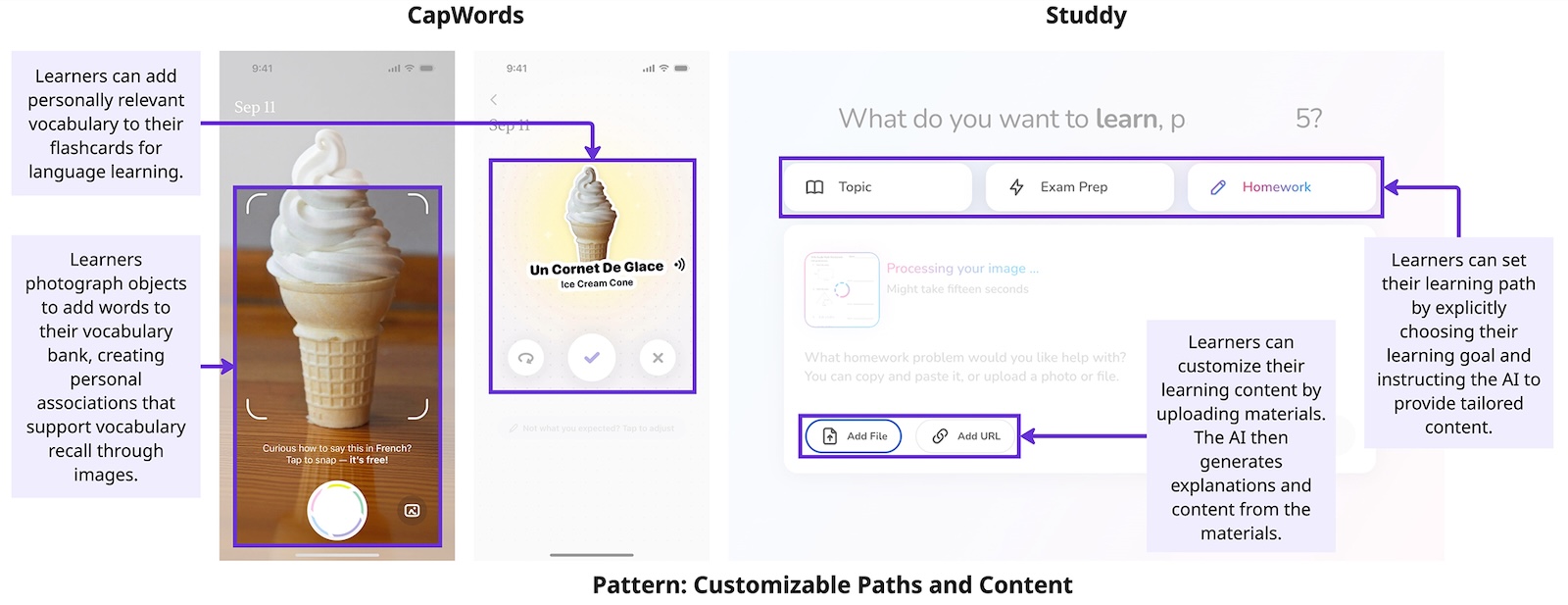}
    \caption{CapWords and Studdy allow learners to personalize their learning by selecting goals, contributing personally meaningful materials, and generating content tailored to their needs.}
    \Description{[DP8.2 Customizable Paths and Content] Screenshots from CapWords and Studdy show two forms of learning personalization. In CapWords, a learner photographs an ice-cream cone and adds the corresponding French term to a personal, image-based vocabulary collection. In Studdy, learners select a goal—topic, exam Prep, or homework—and can upload a file or add a URL. The system then provides guidance and generates learning content based on the selected goal and materials.}
    \label{fig:customizable_paths_and_content}
\end{figure*}

\subsection{Peer Presence}
\label{sec:peer-presence}

\designcommitment{\underline{DC9:} Position peers as comparison targets, audiences, or collaborators according to the form of social participation the learning activity is intended to support.}

Peers can support learning through motivational, cognitive, and social mechanisms. Peer behavior and examples can motivate learners by making the effort and progress of others visible~\cite{bandura1986social}. Peers can
also foster interactive engagement, which is associated with stronger learning outcomes~\cite{chi2014icap}. Collaborative interactions such as explaining reasoning, responding to others, and reaching consensus can deepen
cognitive engagement with learning materials~\cite{tenenbaum2020effective}.

However, in our corpus, the features of the social interface did not create the same kind of peer relationship. \textit{Leaderboards} position peers primarily as comparison targets. \textit{Content Sharing} positions peers as potential audiences for learning materials or work produced by the learner. \textit{Learner Community Support} creates persistent spaces in which peers can become sources of questions, explanations, and alternative approaches. The patterns differ in what other learners become \textit{available for} through the interface.

\begin{table*}
\caption{Patterns associated with the \textit{Peer Presence} design
commitment.}
\label{tab:peer-presence}
\small
\setlength{\tabcolsep}{8pt}
\renewcommand{\arraystretch}{1.30}

\begin{tabularx}{\textwidth}{
    >{\columncolor{PatternLight}}P{0.20\textwidth}
    P{0.37\textwidth}
    Y
}
\toprule
\rowcolor{CommitmentBlue}
\color{white}\textbf{Pattern} &
\color{white}\textbf{Mechanism: Design $\rightarrow$ Learning} &
\color{white}\textbf{Applicability \& Tension} \\
\midrule

\patternname{\underline{DP9.1} Leaderboards}
&
\mechanism{Makes learner activity or performance comparable}
{turning peers into visible reference points for progress, participation, or
competition.}
\corpusexample{Duolingo awards points for practice and displays rankings that
allow learners to compare their activity with peers.}
&
\when{Participation or performance can be represented by a meaningful shared
metric and social comparison is appropriate to the activity.}

\tension{Comparison can provide motivation with little interaction overhead,
but may discourage lower-ranked learners, provide little insight into peers'
learning processes, or shift attention toward point accumulation rather than advancing
understanding.}
\\

\addlinespace[4pt]

\patternname{\underline{DP9.2} Content Sharing}
&
\mechanism{Makes selected learning materials or learner-produced artifacts shareable}
{turning peers into potential audiences and creating opportunities for
discussion around otherwise individual activity.}
\corpusexample{Khan Academy allows individual instructional videos to be
shared; a skill-learning platform allows learners to share images of their
completed projects.}
&
\when{Content or outcomes can be separated into independently meaningful
units that others can understand, respond to, or discuss.}

\tension{Making an artifact shareable does not ensure productive discussion.
Recipients may lack sufficient context, conversations outside the platform may
be difficult to support, and learners may have concerns about sharing work or
data.}
\\

\addlinespace[4pt]

\patternname{\underline{DP9.3} Learner Community Support}
&
\mechanism{Creates persistent spaces for peer questions, responses, and exchange}
{making peers available as ongoing sources of explanation, alternative
approaches, and collaborative support.}
\corpusexample{Udemy allows peers to comment on others' work, and Masterclass and Code Crafters provide pages for discussion.}
&
\when{Large, open-ended, complex, or project-based activities benefit from
sustained exchange and access to diverse approaches.}

\tension{A discussion space does not itself create a functioning community.
Participation requires time and effort, and unanswered questions, inaccurate
advice, uneven participation, and moderation can all limit its usefulness.}
\\

\bottomrule
\end{tabularx}
\end{table*}

These patterns represent qualitatively different forms of peer presence rather than increasing amounts of the same ``social'' feature. A learner who can \textit{see} peers, a learner who can \textit{share with} peers, and a learner who can \textit{work through problems with} peers participate in different social structures. The interface shapes peer learning not simply by making others visible, but by determining what form of relationship and exchange that visibility enables.

\begin{figure*}
    \centering
    \includegraphics[width=\textwidth]{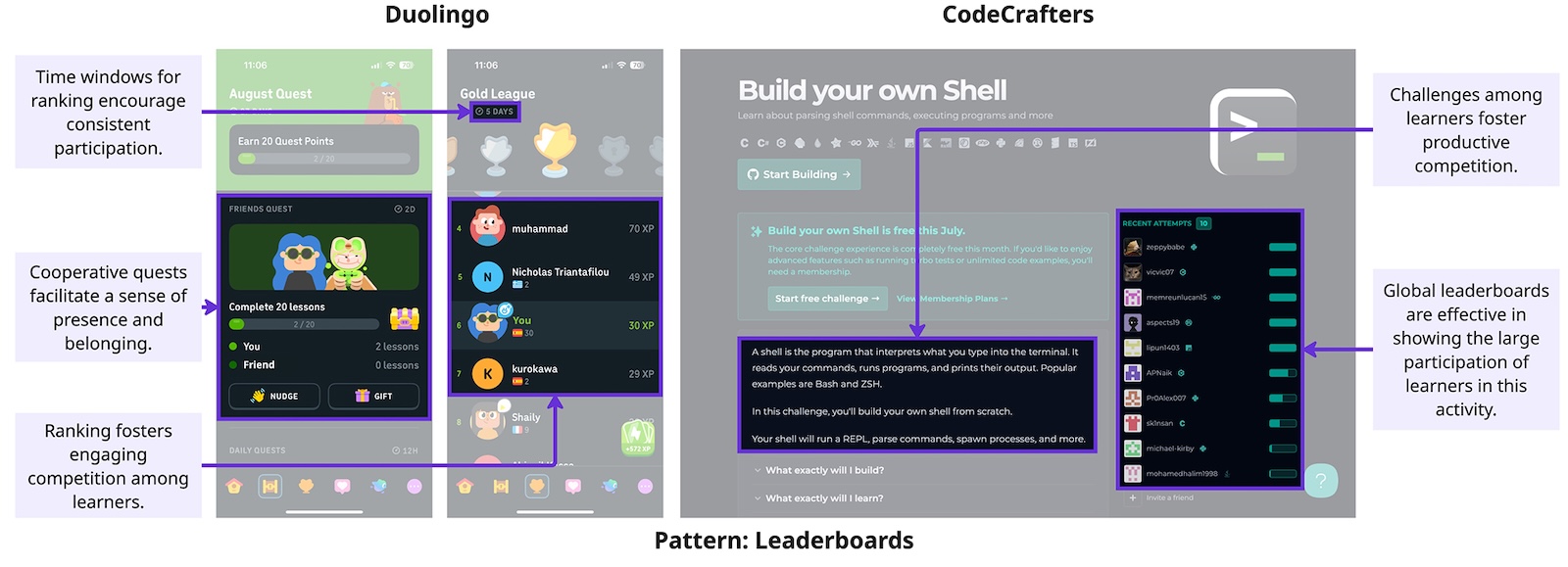}
    \caption{Duolingo and CodeCrafters use quests, rankings, and visible participation indicators to sustain engagement, foster social presence, and encourage productive competition.}
    \Description{[DP9.1 Leaderboards] Screenshots from Duolingo and CodeCrafters demonstrate social and competitive learning features. Duolingo combines a time-limited quest, a cooperative friends quest, and a league ranking that compares learners by experience points. CodeCrafters presents a programming challenge alongside a global list of recent participants and their progress. The examples show how shared challenges, visible participation, and rankings can promote consistency, belonging, and productive competition.}
    \label{fig:leaderboards}
\end{figure*}

\section{Evaluating Pattern Library Use with Designers}
\label{sec:validation}

Following our identification and description of the design patterns, we conducted a qualitative user study with designers. We hypothesized that the patterns surfaced by this study would help designers improve existing learning design interfaces, design new interfaces, and provide a language to think about learning design interfaces. 

\subsubsection{Study Design}
\begin{figure*}
    \centering
    \includegraphics[width=\textwidth]{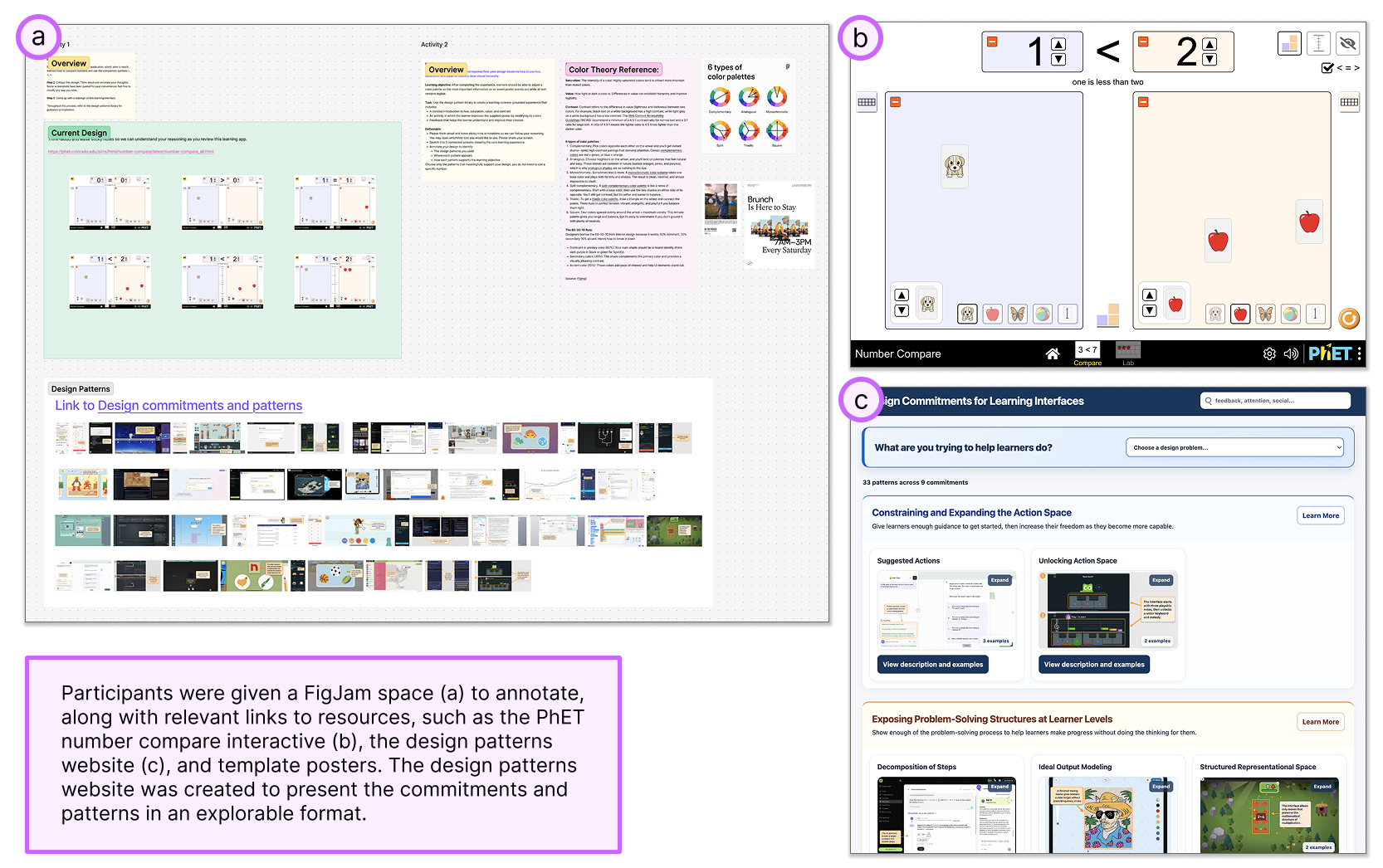}
    \caption{Overview of the materials presented to participants during the user study. Participants were able to explore the design commitments and patterns website before starting activities.}
    \Description{Screenshot of the FigJam space template for participants to annotate, along with relevant links to resources, such as the PhET number compare interactive, the design patterns website, and template posters. The design patterns website was created to present the commitments and patterns in an explorable format.}
    \label{fig:User-study-figjam}
\end{figure*}

To examine how \system might bridge theory and design in practice, we recruited 14 designers in a user study. Each participant was assigned two learning design briefs: first, redesign the existing PhET Number Compare Interactive~\cite{phet}, which was chosen for its simple concepts and served as a warm-up activity. Second, designers were tasked with designing a new learning app to teach first-year design students the basics of color theory. We chose this task because it aligned with participants' domain knowledge and could be completed in a single session. We also provided a "cheat sheet" of color theory~\cite{figma2026colorpalettes} and uncolored poster assets. During both activities, we presented participants with \system in the form of a website, which included text descriptions and annotated screenshots (as shown in Figure~\ref{fig:User-study-figjam}.) This website was created with AI tools, although all text content was written by the authors. During their session, designers were encouraged to think aloud and annotate wireframes to provide rationales for their design choices\cite{wright1991thinkaloud}. All interviews were conducted through Zoom and were recorded and transcribed automatically through Zoom. 

\subsubsection{Participant Recruitment}
We recruited designers by sharing recruitment information on LinkedIn and an alumni-inclusive university email list. At the time of the study, we received 51 responses and recruited 14 who all completed the study. All participants were 18 years or older, based in the U.S., had 2 or more years of design experience, and provided verbal consent to the study and recording. 5 designers did not have prior experience designing for education or learning. The design activity took 90 minutes to complete, and the participants received a \$50 gift card. This study has been approved by our Institutional Review Board.  

\subsubsection{User Study Coding}
We conducted a directed qualitative content analysis of interviews, using categories derived from our study hypotheses and evaluation questions \cite{hsieh2005three, elo2008qualitative}. We coded interview transcripts for instances where participants suggested design improvements, explained their reasoning, described their experience using the library, and commented on the library presentation. During coding, we verified all quotes against video recordings to clarify the referents of the participants.

We then organized coded quotations in a framework matrix, with participants as rows and analytic categories as columns, to support comparisons between participants and design activities \cite{gale2013framework}. We organized quotations by prior work experience, uses of design patterns, other design thinking rationales, experience of using the library, and comments on the library's presentation. 

\begin{figure*}
    \centering
    \includegraphics[width=\textwidth]{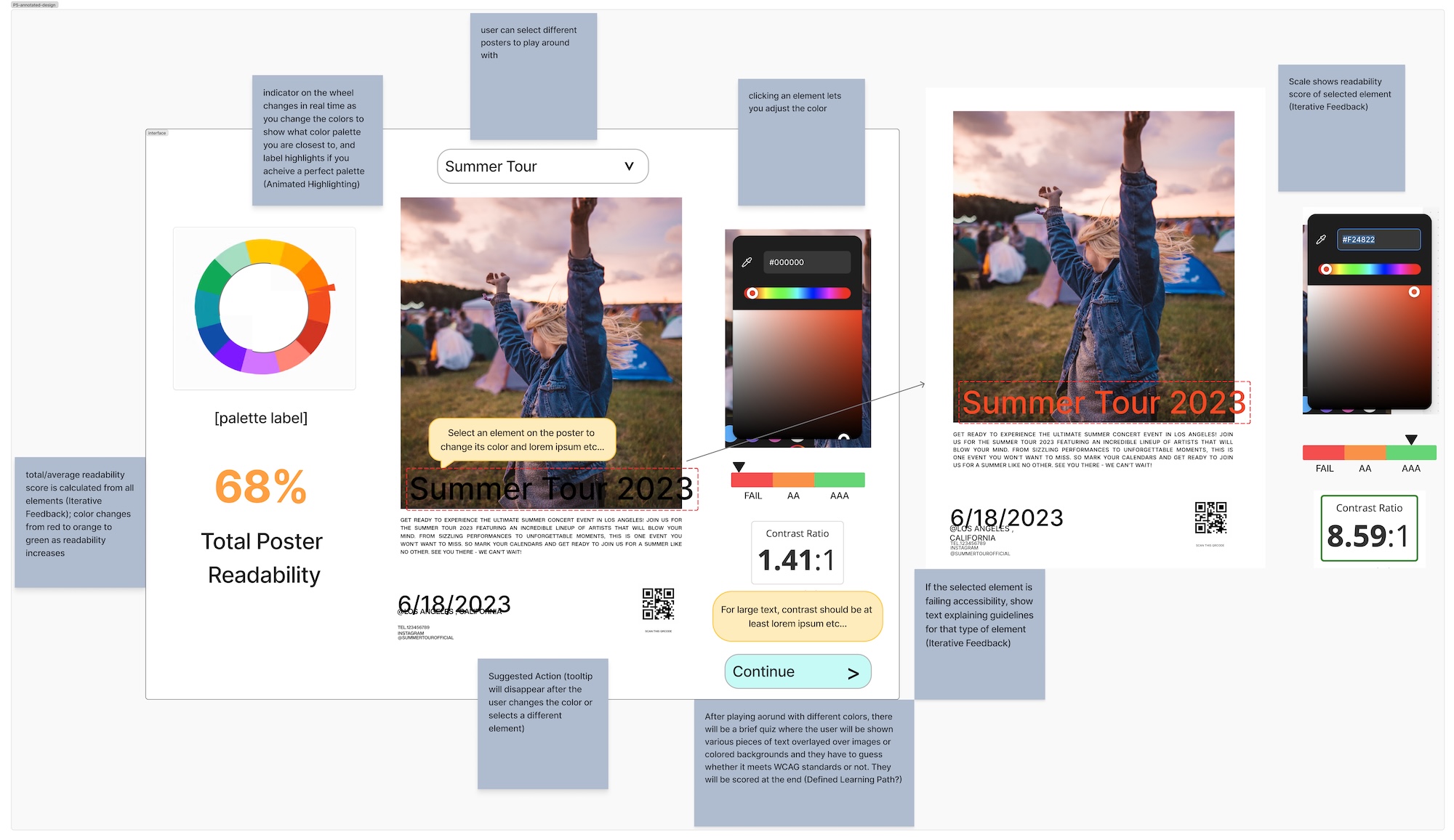}
    \caption{Example from the user study: P5’s annotated interface design for a poster activity that provides feedback on color contrast and readability.}
    \Description{An annotated interface mockup from a study participant that depicts a poster-design activity in which learners select a template and modify element colors using a color wheel and picker. The interface provides iterative accessibility feedback through an overall readability score, element-level contrast ratings, animated highlights, and suggested actions; an example shows the title’s contrast improving from a failing 1.41:1 ratio to an AAA-rated 8.59:1 ratio.}
    \label{fig:Study-example}
\end{figure*}

\subsection{Findings}

For this analysis, we will focus on the second activity, as the redesign activity mainly familiarized designers with the patterns and their application. 13 of the 14 designers selected and implemented design patterns during their design process. Of these, \textit{generative} use ranged from two to ten instances, with a mean of 3.5 and a median of two. We define \textit{generative} \system use as participants explicitly using \system to guide design choices. Four participants applied design patterns without explicitly mentioning them. We interpret the use of unmentioned patterns as evidence that the participants had already internalized these patterns and that the patterns captured existing knowledge of learning-design. In particular, the four designers who used patterns implicitly had prior educational design experience. Lastly, six participants initially misinterpreted or mis-attributed patterns. We attribute these errors to how the patterns were presented, as mentioned below.

\subsubsection{Finding 1: Designers used design patterns to generate interface designs}

The design patterns supported the thinking of the participants both at the outset of the design process and as they developed specific interface features. The participants approached the design pattern library with varying degrees of intentionality. Some used it for open-ended exploration. For example, P4 described \textit{"jumping all around with different things"} before choosing Unlocking Actions and adapting it into an interactive color palette. Others searched for patterns that aligned with specific tasks or learning goals. P3 chose Ideal Output Modeling because \textit{"the example relates more closely to what the task was,"} and reasoned that \textit{"the same concept could apply to the poster making for a student as well."} Other participants recalled design patterns from the introduction and first activity. P9 stated, \textit{"one of the first designs that came to my mind, was the very first design pattern that was... Expanding the space, the action space."} \system guided initial brainstorming from multiple points of entry: open-ended exploration, targeted search, and recall.

Participants also referenced the library to generate a new feature or complement an existing one during their design process. P13 chose Generative Explanations \textit{"since this is color theory, and it relates to arts and design"} to provide learners with individualized feedback. P5 applied Animated Highlighting (Guiding Visual Attention) on a color wheel interactive \textit{"I'd like to use [...] animated highlighting on the indicators on the wheel when it changes."} Similarly, P12 applied Ideal Output Modeling to their existing interface to guide learner actions, \textit{"so that they can understand which output is going to help them more."} Together, these cases illustrate how the library helped participants translate instructional goals into interface elements. 

\subsubsection{Finding 2: The design patterns helped designers justify and communicate their decisions}

During post-activity reflection, participants were asked to reflect on their experience and how design patterns may or may not have influenced their work. Participants indicated that the patterns' learner-centered orientation focused their attention on learner needs and experiences. P2 specifically noted that the design patterns reinforced \textit{"how I should think from the learner's perspective, or understand what is most useful for them. Like, it reinforced that whole concept, otherwise I was just trying to create something [...] with no basis of thinking."} The participants also reflected that \system provided them with a framework to ground and articulate their design decisions. P6 speculated that without \system, they \textit{might have come to a similar redesign,"} but that the design commitments were \textit{"really helpful in helping me explain why,"} suggesting that \system gave participants a conceptual vocabulary to justify their approach.   

\subsubsection{Finding 3: The usefulness of design patterns was limited by information volume, academic language, and reliance on examples}

The participants encountered several challenges in using \system effectively. First, the volume of information made the library difficult to absorb within the time constraints of the study session. P10 explicitly expressed \textit{"I wish I had time,"} to go through the design patterns more thoroughly. Others (P13, P14) mentioned similar difficulty with the amount of information to process within the session's timespan. Their feedback indicates to us that designers would use the patterns more effectively over time and with repeated exposure as they become more familiar with the design patterns. Secondly, the academic terminology used to describe the design commitments and patterns made it difficult for the participants to interpret through text alone. Multiple participants (P4, P7, P8, P12) called the language and titles difficult to understand or vague. Others (P7, P11, P13) explicitly described themselves as visually-oriented. Because the text was harder to understand, some designers relied more heavily on the example images. However, over-reliance on images sometimes resulted in misinterpretations of the pattern. In cases where participants mis-attributed patterns, they were often referencing parts of the example interfaces that were not related to the pattern. These study findings contributed to our last reorganization of design patterns. Because the usefulness of \system was limited by the interpretation of its verbal and visual information, we combined patterns that were difficult to distinguish from each other and revised the organization and descriptions. 

\section{Discussion}
Our findings suggest that the contribution of a pattern library for learning interfaces is to make recurring forms of design reasoning visible, not to prescribe which interface elements designers should use. The library is located at what H\"{o}\"{o}k and L\"{o}wgren call the intermediate level of design knowledge, between the theory of learning on one side and individual systems on the other~\cite{hook2012strong,fang2026frameworks}. The intermediate level connects the interface mechanisms observed across existing learning technologies with broader commitments to learning-related design, preserving the problem, context, and rationale that the form of the pattern is meant to carry~\cite{vanwelie2001patterns,goodyear2005patterns}. Our designer study shows how such a representation is taken up in practice: designers used it to generate and extend interface ideas and, more consistently, to explain why a decision served a learning goal. Their use of the library also confirms that patterns are resources for reasoning as a learning designer rather than recipes for producing effective learning experiences.

Although we present patterns individually, real learning interfaces often instantiate several at once, and these combinations can create competing commitments. The tensions our participants experienced are familiar from the learning sciences: in-context support can reduce incidental cognitive work while removing the effortful processing that produces durable learning~\cite{bjork2011making,gajos2022engage}; explanatory feedback can clarify an error while removing the value of another problem-solving attempt~\cite{kapur2008productive}; and customizable learning paths can increase learner agency while removing the structure upon which learners with limited prior knowledge rely~\cite{kalyuga2003expertise,squires1999usability}. Whether a mechanism helps, therefore, depends on the kind of knowledge being learned and the learner's expertise level   ~\cite{koedinger2012kli}. In terms of patterns, these are the \textit{forces} surrounding a solution~\cite{vanwelie2001patterns}, and they are the reason why the library should not be read as a checklist in which more patterns produce a better design. Its value lies instead in making those forces visible when a designer is choosing a mechanism.

\subsection{Patterns as Design Knowledge for Learning Technology Education}
Across design, computer science, and education, students may encounter courses on designing technologies for learning. However, the knowledge needed to design such systems is distributed across courses and levels of abstraction: learning-science courses introduce theories, while interaction design and computer science courses focus on interfaces, interaction techniques, and implementation. However, the \textit{translation} between them is often overlooked. Students may understand a principle such as scaffolding or metacognitive support while having few examples of the interface mechanisms through which it might be instantiated, or of the tradeoffs those mechanisms introduce, a difficulty consistent with evidence that novices struggle to expand learning goals into concrete design moves~\cite{chang2020novice}. By connecting higher-level design commitments to recurring interface mechanisms, \system makes several such translations visible and comparable. It gives students a repertoire of precedents, which is what experienced practitioners already reason from~\cite{zhou2023lxd,caskurlu2025data}, together with a vocabulary to justify design decisions in terms of both the learner and the user~\cite{schmidthuang2022lxd,soloway1994learner}.

\subsection{Representing Design Knowledge for Use}\label{sec:disc-representation}
Designers' difficulties with terminology, volume, and examples point to a design problem in the representation itself. Pattern names are meant to become a shared vocabulary~\cite{alexander1977language,borchers2000patterns}, yet several of ours had to be decoded before they could be used. Participants also worked primarily on the specifics of the annotated examples, sometimes reproducing their surface form while missing the mechanism they illustrated, a fixation effect well documented in design ideation~\cite{jansson1991fixation}. Examples are what make a pattern usable, but they also make its rationale easy to bypass, which is the failure mode that Dearden and Finlay describe when the pattern form is adopted without taking on its underlying values~\cite{dearden2006patterns}. Developing a pattern library is therefore both an interaction design problem and an authoring one.

\subsection{Patterns for Generative and Bespoke Learning Software}
The need for this kind of intermediate design knowledge is likely to become more vital, as generative AI lowers the cost of producing interactive software. Learners, teachers, designers, and developers can increasingly generate small applications, simulations, tutors, and interfaces for highly specific learning goals rather than relying only on professionally designed, general-purpose educational platforms. In such settings, a pattern library can provide conceptual scaffolding for this emerging form of software creation. A person prompting a generative interface system could, for example, reason explicitly about whether a task should expose intermediate problem structure, what information should remain for the learner to retrieve, how feedback should preserve opportunities to retry, or who should determine the progression of content. Patterns could therefore inform not only manual design practice but also the prompts, specifications, critiques, or iterative conversations through which bespoke learning applications are generated.

This possibility also creates a risk. Generative systems make it easy to reproduce recognizable interface forms without preserving the learning rationale that made them appropriate in their particular context. A generated ``leaderboard,'' ``hint,'' or ``worked example'' may look plausible while being poorly aligned with the learner, domain, or intended learning activity. The usefulness of the patterns in generative workflows therefore lies less in providing a menu of features to request, and more in making the conditions and tensions surrounding an interface mechanism explicit for consideration and warranted disputation. Applicable to our context, Argyris (1996) \cite{argyris1996organizational} distinguished Single-Loop vs. Double-Loop Learning: When educators (learning designers in our case) merely perform the surface features of an innovation without shifting their mental models, they engage in single-loop learning (changing tactics to meet an outward metric). True innovation requires double-loop learning, where practitioners fundamentally question and understand the underlying rationale and values of the new model.

\subsection{Limitations and Future Work}
Our patterns describe recurring mechanisms in existing learning interfaces; they do not establish that these mechanisms improve learning. Recurrence is evidence that a design problem and a response appear across systems, not that the response is optimal for a particular learner, domain, or context. The learning rationales we associate with the patterns are analytic interpretations rather than evidence of the intentions of the original designers, closer to design conjectures that remain to be tested than to established mappings~\cite{sandoval2004conjectures}.

The corpus also reflects the applications that we were able to identify and access. Although we intentionally sampled across domains and interface forms, the corpus is neither exhaustive nor representative of all learning technologies: paid, inaccessible, emerging, culturally specific, or less widely-distributed systems may contain mechanisms not captured here. Furthermore, the preponderance of the learning applications we identified focused on the individual learner, so the vibrant research field of CSCL (Computer-Supported Collaborative Learning), with its extensive literature on 'collaborative learning scripts' \cite{dillenbourg2007flexibility, kollar2006collaboration, vogel2021collaboration} is not well-represented in the screenshots nor, accordingly, the Design Commitments and Design Patterns. Remedying this CSCL omission deserves its own research project. Our designer study also examines bounded design activities; it does not demonstrate that the use of patterns leads to better interfaces or improved learning outcomes.

The library also does not include accessibility as a design commitment. Across the corpus, the interfaces addressed differences in visual ability, reading ability, proficiency in the interface language, and sensitivity to visual clutter through mechanisms such as audio descriptions, alternative text for images, language and reading support, and controls that allow learners to show, hide, or modify visual information. These mechanisms cross commitments rather than serve a particular form of learning activity: toggling graphics alters visual attention and cognitive load, alternative representations change how information is accessed, and language or reading support can determine whether a learner can participate in the focal task at all. We chose not to reduce them to a single pattern or commitment because accessibility is not an optional feature that can be added to an otherwise complete design. Presentation controls expand access and let learners adapt aspects of an experience to their needs, but they do not substitute for making the underlying content, representations, and interactions accessible by design. Whether accessibility warrants design commitments of its own, and how each of the present commitments should be realized for learners with different competencies are questions the library leaves open.

These limitations motivate several directions for future work: individual patterns and their combinations can be evaluated in relation to learner behavior and learning outcomes, CSCL design commitments and design patterns should be investigated, and future studies can examine how the library is used in longer design processes and by multidisciplinary teams. Generative and adaptive interfaces, in particular, may assemble combinations of mechanisms dynamically rather than committing to a single fixed interface, so that decisions about when to introduce and fade support~\cite{quintana2004scaffolding} are made by the system during use rather than by a designer in advance. How design commitments should operate in such systems is an open question for learning-interface research.


\section{Conclusion}


Learning interfaces embody designers' decisions about how learning theory becomes a force shaping what a learner sees and does in their learning environment, but the reasoning behind those decisions is seldom documented or explicitly taught. Designers make such decisions by drawing on precedent and experience, and the knowledge stays within the products. In this paper, we analyzed 873 screenshots from 65 learning applications and identified 27 recurring interface patterns, organized under nine design commitments that connect each pattern to the learning concern it addresses. We then observed 14 designers working with the library on two design tasks. They used the patterns to come up with interface ideas and, more often, to explain why a given design choice served a specific learning goal. Their feedback also demonstrated what makes such a library usable in practice: plain names, a manageable number of patterns, and examples that carry their rationale with them rather than standing in for it. Our patterns document what designers currently do; whether these mechanisms actually improve learning, alone or in combination, remains to be tested. We also see \system as a first step toward a pattern language for learning interfaces, one that would describe not only individual mechanisms but also how they combine, so that learning theory and design practice can inform each other more directly, in principled ways.

\bibliographystyle{ACM-Reference-Format}
\bibliography{99_refs}

\end{document}